\documentclass[fleqn,usenatbib]{mnras}
\usepackage{newtxtext,newtxmath}
\usepackage[T1]{fontenc}
\DeclareRobustCommand{\VAN}[3]{#2}
\let\VANthebibliography\thebibliography
\def\thebibliography{\DeclareRobustCommand{\VAN}[3]{##3}\VANthebibliography}

\usepackage{graphicx}	
\usepackage{amsmath}	
\usepackage{subcaption}
\usepackage{algorithm}
\usepackage{algpseudocode}
\usepackage[normalem]{ulem}

\title[Geometry of $k$NNs]{Geometric Interpretations of the $k$-Nearest Neighbour Distributions}
\author[K. Gangopadhyay, A. Banerjee and T. Abel]{
Kwanit Gangopadhyay,$^{1,2}$\thanks{E-mail: k.gangopadhyay@rug.nl}
Arka Banerjee,$^{2}$\thanks{E-mail: arka@iiserpune.ac.in}
Tom Abel$^{3,4,5}$
\\
$^{1}$ Van Swinderen Institute for Particle Physics and Gravity, University of Groningen, Nijenborgh 3, 9747 AG Groningen, The Netherlands \\
$^{2}$Department of Physics, Indian Institute of Science Education and Research, Pune 411008, India\\
$^{3}$Department of Physics, Stanford University, 382 Via Pueblo Mall, Stanford, CA 94305, USA\\
$^{4}$Kavli Institute for Particle Astrophysics \& Cosmology, P. O. Box 2450, Stanford University, Stanford, CA 94305, USA\\
$^{5}$SLAC National Accelerator Laboratory, Menlo Park, CA 94025, USA
}
\date{Accepted XXX. Received YYY; in original form ZZZ}
\pubyear{2025}

\begin{document}
\label{firstpage}
\pagerange{\pageref{firstpage}--\pageref{lastpage}}
\maketitle

\begin{abstract}
The $k$-Nearest Neighbour Cumulative Distribution Functions are measures of clustering for discrete datasets that are fast and efficient to compute. They are significantly more informative than the 2-point correlation function. Their connection to $N$-point correlation functions, void probability functions and Counts-in-Cells is known. However, the connections between the CDFs and geometric and topological summary statistics are yet to be fully explored.This understanding will be crucial to find optimally informative summary statistics to analyse data from stage-4 cosmological surveys. We explore quantitatively the geometric interpretations of the $k$NN CDF summary statistics. We establish an equivalence between the 1NN CDF at radius $r$ and the volume of spheres with the same radius around data points. We show that higher $k$NN CDFs represent the volumes of intersections of $\ge k$ spheres around data points. We present similar geometric interpretations for the $k$NN cross-correlation CDFs. We further show that the full shape of the CDFs have information about planar angles, solid angles and arc lengths created at the intersections of spheres around the data points, and can be accessed through the derivatives of the CDF. We show that this information is equivalent to that captured by Germ Grain Minkowski Functionals. Using Fisher analyses we compare the constraining power of various data vectors constructed from $k$NN CDFs and Minkowski Functionals. We find that the CDFs and their derivatives and the Minkowski Functionals have nearly identical constraining power. However, the CDFs are computationally orders of magnitude faster to evaluate. 
\end{abstract}

\begin{keywords}
methods: statistical - large-scale structure of Universe - cosmological parameters 
\end{keywords}


\section{Introduction}
One of the primary objectives of cosmology has been to verify the $\Lambda$CDM model of the universe and constrain the cosmological parameters, using a variety of cosmological probes, ranging from the cosmic microwave background to the large scale structures in the late universe. These studies require extracting information from cosmological fields - either continuous, such as the CMB field (e.g., \cite{refId0}) and weak lensing shear maps (e.g., \cite{PhysRevD.98.043528}, \cite{Dalal_2023}), or discrete, such as galaxy positions (e.g., \cite{Abbott_2022}) and galaxy cluster positions (e.g., \cite{2021AAS...23750103P}). Over the past few decades, there have been several efforts at constraining the cosmological models using these observations. 
\\ \\
The matter field in the early universe was nearly a Gaussian Random Field (\cite{Peacock_1998}, \cite{refId3}). However, over time, gravitational collapse led to clustering of matter. This evolution of the matter field is well described by linear perturbation theory on sufficiently large scales ($\gtrsim$50 Mpc/h) even at $z=0$. Under linear evolution, a Gaussian random field remains Gaussian (\cite{Baumann_2022}), with only the variance as a function of scale evolving with time (\cite{2012arXiv1208.5931K}). Such a field can be completely summarized by the power spectrum, or the 2-point correlation function (\cite{2012arXiv1208.5931K}). At smaller scales at late times, linear perturbation theory breaks down (\cite{2020moco.book.....D}). The matter field is non-Gaussian at these scales, forming a highly clustered web-like structure. In this regime, the 2-point function becomes an incomplete statistic, failing to capture a significant portion of the information related to the non-linear collapse of matter.
\\ \\
Much of the analysis on observational data over the past few decades has been using the 2-point correlation function or the power spectrum. This includes landmark achievements such as the precise measurement of the CMB power spectrum (\cite{refId2}) and the detection of Baryon Acoustic Oscillations in the galaxy two-point correlation function (\cite{Eisenstein_2005}). These observations have helped refine our cosmological models and establish $\Lambda$CDM as a fiducial model of the universe. But with increasing sensitivity of detectors, surveys such as DESI, Euclid, and LSST are expected to deliver unprecedented volumes of data at non-linear scales. Consequently, more effective methods are required to extract valuable information from these scales. This in turn will help refine our cosmological models further.
\\ \\
Several higher-order statistics and techniques have been proposed to address this challenge by leveraging the clustering of discrete tracers: the $N$-point correlation functions and their Fourier transforms (e.g., \cite{1975ApJ...196....1P}, \cite{1978ApJ...221...19F}, \cite{10.1093/mnras/186.2.145}, \cite{10.1111/j.1365-2966.2004.07410.x}, \cite{PhysRevD.74.023522}, \cite{10.1093/mnras/stv961}), Void Probability Function (e.g., \cite{1986ApJ...306..358F}, \cite{10.1093/mnras/stz1351}), Counts-in-Cells (e.g., \cite{1994ApJ...420...44K}, \cite{PhysRevD.90.103519}, \cite{10.1093/mnras/sty2613}), Minkowski Functionals (e.g., \cite{1994A&A...288..697M}, \cite{Schmalzing:1995qn},\cite{10.1093/mnras/stu1118}), Betti Numbers (e.g., \cite{10.1093/mnras/stw2862}, \cite{vandeWeygaert2011}, \cite{10.1093/mnras/stad1765}), Voronoi Volume Functions (\cite{10.1093/mnras/staa1379}), and Persistent Homology (e.g., \cite{Biagetti_2021}, \cite{2023arXiv230802636Y}, \cite{Yip_2024}) being some of them. These techniques capture additional information from the clustering of tracers such as galaxies, at small, non-linear scales where the 2-point function proves inadequate. While some of these statistics, such as the Void Probability Function (VPF) and Counts-in-Cells (CICs), are applicable only to discrete datasets, others, like the Minkowski Functionals, can also be applied to continuous fields. For example, Minkowski Functionals have been used to analyze the non-Gaussian thermal Sunyaev-Zel’dovich map from the CMB (\cite{2024arXiv241021247S}) and the H\;\textsc{i} field around dark matter halos (\cite{10.1093/mnras/stab1555}). Some of them can also be extended to study cross-correlation between different datasets ($N$-point functions, e.g., \cite{1974ApJS...28...37P}, \cite{1980ApJ...238..785F}) while others might not be suited for it (Minkowski Functionals). Beyond spatial clustering, number counts or mass functions of halos, galaxies, clusters, and other objects (e.g., \cite{Allen:2011zs}, \cite{Tinker_2012}) can also be employed to constrain cosmological models.
\\ \\
For these summary statistics to be useful for large surveys, they should be fast, robust, informative, and interpretable. Being computationally fast and efficient will enable these statistics to be applied to large datasets from upcoming surveys. This approach will accelerate the analysis process and enhance cosmological constraints by enabling the efficient utilization of larger datasets. They should also respond roughly linearly to systematic errors, and we should be able to understand and predict these offsets to correctly callibrate our statistics. They should be able to detect clustering even in noisy data and be sensitive to small changes in cosmological parameters. Finally, these statistics should be interpretable, allowing us to identify which aspects of the underlying clustering drive their information content. By understanding how they relate to other statistics, we can effectively combine multiple approaches, maximizing the extraction of information and achieving the tightest possible constraints. All the higher-order statistics mentioned above are more informative than the 2-point correlation function. But some of them (like Persistent Homology or $N$-point functions) are computationally very expensive in existing implementations.
\\ \\
The $k$-Nearest Neighbour Cumulative Distribution Functions are a set of summary statistics introduced in \cite{10.1093/mnras/staa3604}. They have been shown to be more informative than the 2-point correlation functions, sensitive to all the $N$-point correlation functions, and computationally inexpensive to evaluate. In addition, the $k$NN CDFs can be used to measure clustering in discrete data (\cite{10.1093/mnras/staa3604}, \cite{Wang:2021kbq}), continuous data (\cite{DES:2023juo}), cross-correlation between two discrete datasets (\cite{10.1093/mnras/stab961}) and cross-correlation between a discrete dataset and a continuous field (\cite{10.1093/mnras/stac3813}, \cite{Gupta:2024sss}, \cite{Zhou:2024kzp}, \cite{Chand:2024vpg}). The CDFs can also be modelled using a hybrid effective field theory approach (\cite{Banerjee:2021cmi}), allowing us to further understand the tracer-matter connection and use these statistics on real observational data from cosmological surveys. Furthermore, disentangling the effects of redshift-space distortions by decomposing the $k$NN CDFs into components parallel and perpendicular to the line-of-sight leads to even stronger parameter constraints (\cite{10.1093/mnras/stad1275}).
\\ \\
Understanding the CDF in different mathematical languages by exploring its connections with other summary statistics will help us understand its systematics and use a combination of various statistics to obtain the best possible constraints on cosmological parameters. The connections between the $k$NN CDFs and other statistics, such as 
$N$-point functions, Counts-in-Cells (CICs), and the Void Probability Function (VPF), have been established (\cite{10.1093/mnras/staa3604}). However, the relationships between $k$NN CDFs and other statistical measures remain less explored. Unlocking these connections could provide new insights and interpretations for these statistics, linking them in novel and meaningful ways.
\\ \\
In this paper, we extend the interpretability of these distributions and identify systematic effects that influence their practical application in constraining cosmological parameters from data. We introduce a geometric interpretation of the 1NN CDF in terms of volume within spheres around the discrete tracers. We extend this interpretation to explain $k$NN CDFs for $k>1$, as well as cross-correlation joint CDFs. We then present the geometric information contained in the derivatives of these CDFs, demonstrating that their interpretations align closely with those of the Minkowski Functionals.  We show a correspondence between the 1NN CDF and its derivatives and the Minkowski Functionals, and, through a Fisher analysis, find that the CDFs and derivatives and the Minkowski Functionals provide nearly identical constraints on the cosmological parameters. Thus, with this work, we have established a common framework connecting multiple higher-order statistics describing both the algebraic and geometric properties of the cosmological fields. Finally, we identify potential improvements in the construction of our data vector and emphasize the need for more efficient compression methods for the CDFs to achieve tighter constraints.
\\ \\
The rest of the paper is arranged as follows: in section \ref{sec:formalism}, we discuss the mathematical framework describing the $k$NN CDFs, their geometric interpretations and their connection with various other higher-order statistics. We discuss in detail the connection between the 1NN CDF and the Minkowski Functionals in section \ref{sec:knn_mink}. In section \ref{sec:fisher}, we explain the methodology and results of our Fisher analysis, and in section \ref{sec:final}, we summarize this work and discuss the key takeaways and future directions. 

\section{Mathematical Formalism}
\label{sec:formalism}
The $k$NN CDFs as a measure of clustering was introduced in \cite{10.1093/mnras/staa3604}. They have been demonstrated to be highly sensitive to non-linear clustering information on quasi-linear to non-linear scales. The $k$NN CDFs were already shown to be sensitive to all the higher-order $N$-point correlation functions, as well as the void probability function and the counts-in-cells. Here, we establish the geometric interpretations of these statistics.

\subsection{$k$-Nearest Neighbour Distributions}
We start with a set of discrete data points spread over space. In this paper, we focus on 3-dimensional datasets, though many of our results can also be applied to 2-dimensional datasets, which are also relevant for cosmological surveys (\cite{Gupta:2024sss}). The $k$-Nearest Neighbour Cumulative Distribution Function for a length $r$ is simply the volume averaged probability of finding greater than or equal to $k$ data points within randomly drawn spheres of radius $r$ in the space: 
\begin{equation}
    \mathrm{CDF}_{k\mathrm{NN}} (r) = P_{\ge k | V = \frac{4}{3}\pi r^3}
    \label{eq:definition}
\end{equation}
We use the volume of the sphere $V$ and its radius $r$ interchangeably ($V = \frac{4}{3}\pi r^3$). $P_{\ge k|V}$ represents the fraction of points in space for which a sphere of radius $r$ centered around those points contain at least $k$ data points. On the other hand, $\mathrm{CDF}_{k\mathrm{NN}} (r)$ represents the fraction of points in space from which the $k^\textrm{th}$ nearest neighbour data point is at a distance of at most $r$, $P(k\mathrm{NN\hspace{0.1cm}distance}\leq r)$. These two definitions are equivalent, since the points that have $\ge k$ data points within a distance $r$ are the same points from which the distance to the $k^\textrm{th}$ nearest neighbour data point is $\le r$. If a point has exactly $k$ data points within a distance $r$, then its $k^\textrm{th}$ nearest neighbour data point must be within a distance $r$ and the $k+1^\textrm{th}$ nearest neighbour data point must be at a distance $\ge r$. This point is then considered while computing all the CDFs till $\mathrm{CDF}_{k\mathrm{NN}}$, and is not considered for computing $\mathrm{CDF}_{(k+\alpha) \mathrm{NN}}$ for positive integers $\alpha$.
\\ \\
To empirically calculate this quantity in real data, we randomly sample a dense set of points in the space (call them \textit{query} points), instead of drawing the infinitely possible spheres for each radius $r$. Note that we can even choose to create a uniform grid of densely spaced query points instead of random sampling. 
\newline \newline \textbf{Step 1:} We first fill our space with Poisson distributed query points. Their number density should ideally be higher than the number density of the data points. Alternatively, we can place the query points on a grid if the grid's separation is much smaller than the mean distance between the data points. 
\newline \newline \textbf{Step 2:} From each query point, we calculate the distances to the nearest neighbour data point ($k=1$), the second nearest neighbour data point ($k=2$), and so on. Thus, for each $k$ we get a list of as many distances as the number of query points.
\newline \newline \textbf{Step 3:} For a given $k$, we sort the list of the $k$NN distances. In the ascending list of distances, the $i^{\textrm{th}}$ distance (say, $r_{i}$) gives us the following relation for the empirical CDF:
\begin{equation}
        \mathrm{CDF}_{k\mathrm{NN}} (r_{i}) = \frac{i}{N_{q}}
\end{equation}
where $N_q$ is the total number of query points. We can interpolate this discretely defined function to get a smooth function. The smooth function will be our approximation to the true $\mathrm{CDF}_{k\mathrm{NN}}$.
\\ \\
In this algorithm to compute the CDFs, Step 2 can be done in a quick and efficient way ($ N\log N$) using a $k$-d tree (\cite{4061547}), which has been implemented in the \textsc{cKDTree} implementation of \textsc{SciPy}\footnote{\url{https://docs.scipy.org/doc/scipy-1.15.0/reference/generated/scipy.spatial.cKDTree.html}}. If we have $N_d$ data points and $N_q$ random query points, then the list of distances can be computed in $N_q \log N_d$ operations. Thus, the $k$NN CDFs serve as a quick, efficient and effective summary statistic for cosmological clustering at non-linear scales. 

\subsection{$k$NN CDFs and Counts-in-Cell statistics}
Here, we will briefly summarize the results already obtained in \cite{10.1093/mnras/staa3604}, exploring the connections between the $k$NN CDFs and the CICs. The counts-in-cell statistic measures the probability of finding exactly $k$ data points within randomly drawn spheres of volume $V$ in the space ($P_{k|V}$). The CIC and $k$NN CDFs are related by:
\begin{equation}
\begin{aligned}
    P_{k|V} &= P_{\ge k|V} - P_{\ge k+1|V} \hspace{1cm} \forall k \ge 0  \\
     &= \mathrm{CDF}_{k\mathrm{NN}}(r) - \mathrm{CDF}_{(k+1)\mathrm{NN}}(r)\hspace{0.5mm}; \hspace{0.2cm}  \text{ where } \frac{4}{3}\pi r^3 = V
\end{aligned}
    \label{eq:CIC_def}
\end{equation}
For $k=0$, $P_{0|V}$ is called the Void Probability Function (VPF) (\cite{10.1093/mnras/186.2.145}). It is simply the probability of finding no data points within randomly drawn spheres of volume $V$ in the space. Since $P_{\ge 0|V} = 1$ and $P_{\ge 1|V} = \mathrm{CDF}_{1\mathrm{NN}}$, using equation~\ref{eq:CIC_def} we can write the VPF as
\begin{equation}
    \mathrm{VPF}(r) = 1 - \mathrm{CDF}_{1\mathrm{NN}}(r)
    \label{eq:vpf}
\end{equation}
The VPF has been used as a summary statistic where it is used both as a function of the radius $r$ and the mean number density of particles $\bar{n}$. Since equation \ref{eq:vpf} connects the VPF and the 1NN CDF at some given $r$, both contain the same information for a constant $\bar{n}$. The VPF is traditionally computed by randomly drawing numerous spheres of radius $r$ and counting what fraction do not have a data point. This algorithm requires to be run for every radius $r$ we are interested in. However, the $k$NN CDFs are computed using the $k$-d tree algorithm and by sorting the distances from query points just once, resulting in it being much faster to compute than the VPF.
\\ \\
Let $P(z|V)$ and $C(z|V)$ be the generating functions of the CICs and $k$NN CDFs respectively. The $k$NN CDF can alternatively be written as $P_{> k-1|V}$ instead of $P_{\ge k|V}$. Then,
\begin{equation}
    P(z|V) = \sum_{k=0}^\infty P_{k|V}\, z^k \; , \hspace{5mm}
    C(z|V) = \sum_{k=0}^\infty P_{>k|V}\, z^k
\end{equation}
Then, it can be shown that (see section 2.1 of \cite{10.1093/mnras/staa3604}) 
\begin{equation}
    C(z|V) = \frac{1 - P(z|V)}{1 - z}
    \label{eq:CIC_kNN}
\end{equation}
Thus, if we hold the mean number density of particles $\bar{n}$ constant, the counts-in-cells and nearest neighbour distributions are directly related to each other. But computing the the nearest neighbour distributions by sorting distances from query points using a $k$-d tree algorithm is much faster than computing the CICs by sampling numerous random spheres and counting the number of data points within them. 

\subsection{$k$NN CDFs and $N$-point correlation functions}
Here, we summarize the connection between the $N$-point functions and the $k$NN CDFs. A detailed discussion and derivation can be found in section 2.1 and Appendix A of \cite{10.1093/mnras/staa3604}). 
\\ \\
For a discrete set of data points with a mean number density of $\Bar{n}$ tracing an underlying smooth field, the generating function of the CICs can be written as (\cite{1993ApJ...408...43S}) 
\begin{align}
P(z|V) &= \exp \left[ \sum_{k=1}^{\infty} \frac{\bar{n}^k (z - 1)^k}{k!} \; \times \right. \nonumber \\
       &\quad \left.  \int_V \dots \int_V d^3\mathbf{r}_1 \dots d^3\mathbf{r}_k \; \xi^{(k)}(\mathbf{r}_1, \dots, \mathbf{r}_k) \right]
\end{align}
where $\xi^(N)((\mathbf{r}_1, \dots, \mathbf{r}_N))$ is the connected $N$-point correlation function. $\xi^(0) = 0$ by definition, and we choose $\xi^(1) = 1$ for normalization. Note that $P(z|V)$ is related to $C(z|V)$, the generating function of the $k$NN CDFs, by equation \ref{eq:CIC_kNN}. Thus, the $k$NN CDFs are sensitive to all the $N$-point correlation functions. But computing the $k$NN CDFs is much faster than computing the correlation functions. 

\subsection{The geometric interpretations of $k$NN CDFs}
\label{sec:knn_geo}

\begin{figure}
    \centering
    \includegraphics[width=\linewidth]{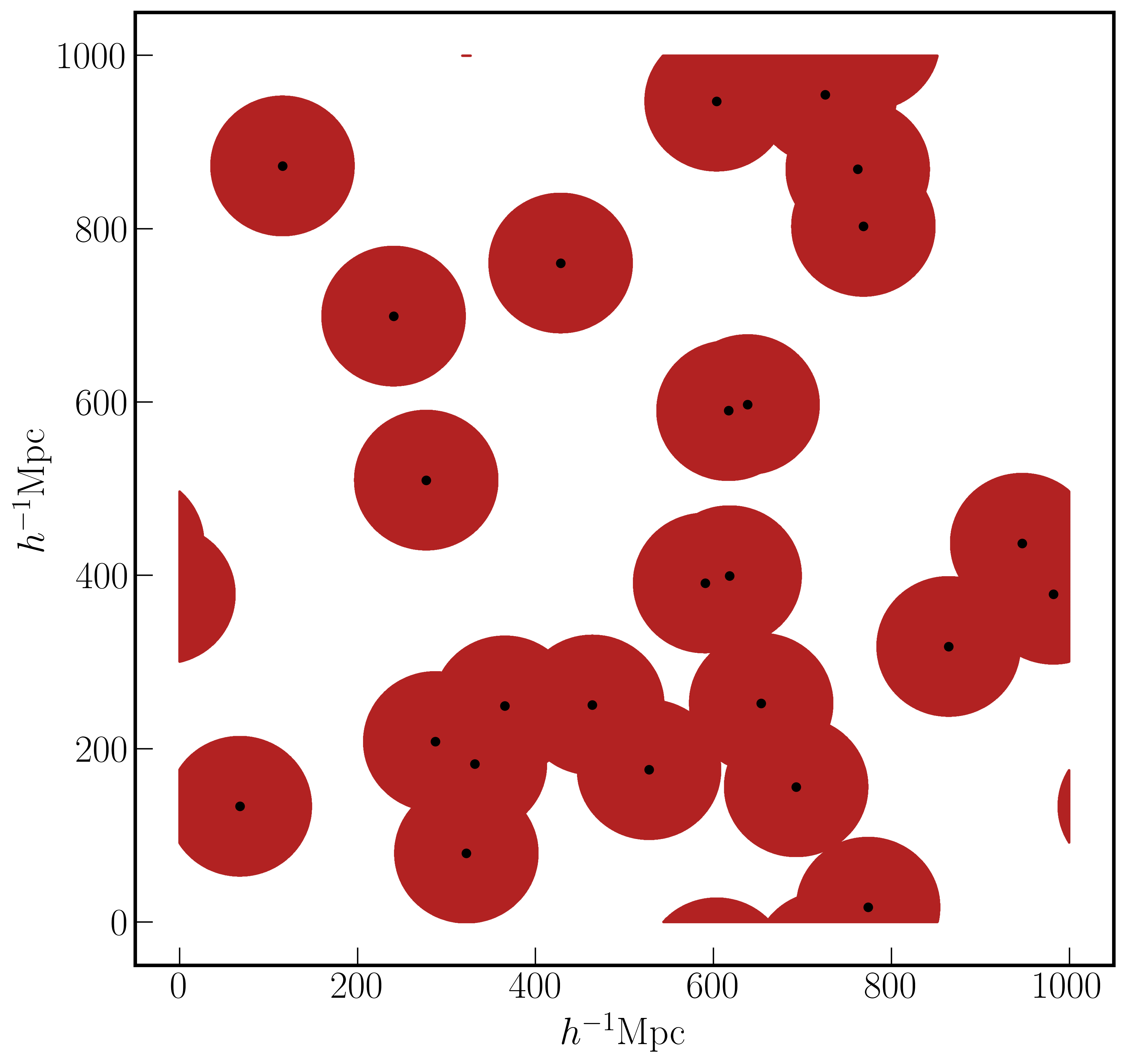}
    \caption{Shaded are all the query points that have a 1NN distance of $\le r$ from the data points (in black). Thus, the fraction of the total box that is shaded gives us the $\mathrm{CDF}_{1\mathrm{NN}}$ at this particular radius $r$ (equation~\ref{eq:W0_1}). This is a representative figure with circles in 2D, but the same argument holds with spheres in 3D. }
    \label{fig:1nn_geo}
\end{figure}

The $\mathrm{CDF}_{1\mathrm{NN}}$ is a measure of the fraction of (query) points in space are within a distance $r$ of a data point (hence having a 1NN distance $\le r$). The set of all points in space that are a distance of at most $r$ from the data points forms spheres of radius $r$ around data points. Note that we have gone from drawing infinite spheres (one around every point in space) and counting the number of data points within these spheres to simply drawing spheres around the finitely many data points and calculating the volume within this union of spheres (see figure \ref{fig:1nn_geo}). If we are dealing with a simulation box with total volume $V_0$ then the volume $V(r)$ within spheres of radius $r$ drawn around data points is 
\begin{equation}
    V(r) = V_0\times \mathrm{CDF}_{1\mathrm{NN}}(r)
    \label{eq:W0_1}
\end{equation}
\begin{figure}
    \centering
    \includegraphics[width=\linewidth]{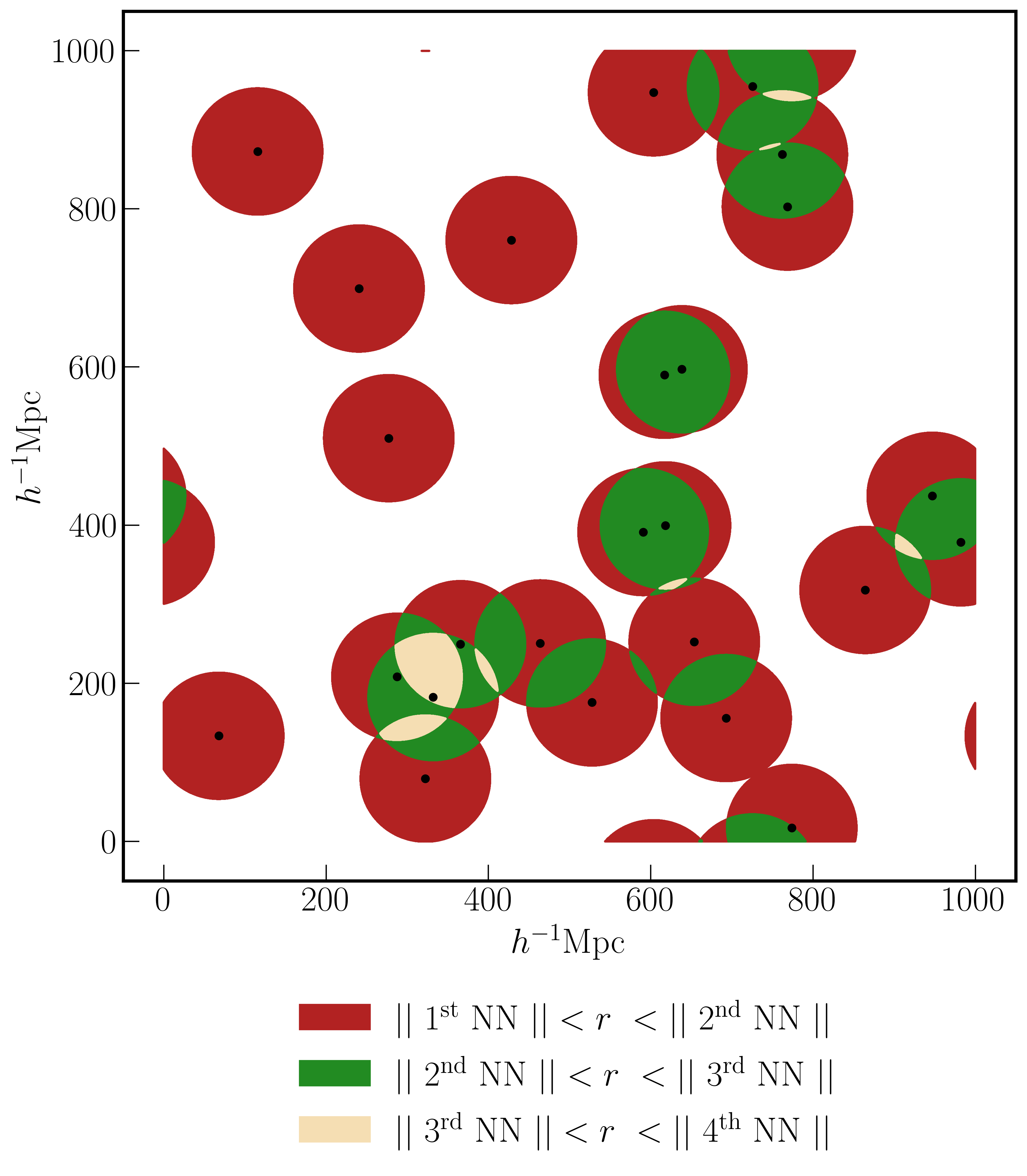}
    \caption{Shaded in different colours are the volume within exactly one sphere, exactly two spheres and exactly three spheres. These volumes can be obtained by using the $\mathrm{CDF}_{k\mathrm{NN}}$'s using equation \ref{eq:difference_of_cdfs}. The symbol $||.||$ denotes the Euclidean distance, so $|| \; 1^{\mathrm{st}}\; \textrm{NN}\; || < r$ means that distance to the first nearest neighbour is less than $r$. The $k$NN CDF corresponds to the fraction of volume within \textit{at least} $k$ spheres, so it is the sum of the volumes within intersections of exactly $k, k+1, k+2, ...$ spheres. Hence, the 1NN CDF is the proportional to the sum of volumes of the red, green and beige regions; the 2NN CDF proportional to the sum of volumes of the green and beige regions, and the 3NN CDF proportional to the volume of the beige region.}
    \label{fig:knn_geometric}
\end{figure}
\\
\noindent 
This shift in perspective from drawing spheres around query points to drawing spheres around data points is going to greatly help us devise several geometric interpretations related to the 1NN CDF, as well as relate it with the Minkowski Functionals (section \ref{sec:knn_mink}). \\ \\
Points within the intersection region of two or more spheres of radius $r$ around data points have both the 1NN and 2NN distance $\le r$, hence being within radius $r$ of those data points. Thus, the volume created by intersection of two or more spheres is given by $V_0\times \mathrm{CDF}_{2\mathrm{NN}}(r)$. Similarly, points within the intersection region of three or more spheres of radius $r$ centered around data points have their 1NN, 2NN and 3NN distance $\le r$. Thus, the volume created by intersection of three or more spheres is given by $V_0\times \mathrm{CDF}_{3\mathrm{NN}}(r)$, and so on for higher $k$NN CDFs. If we have a total of $N_d$ data points and draw spheres of radius $r$ around them, each sphere $S_i$ occupying a volume of $\frac{4}{3}\pi r^3$ around the data point $i$, then the volume within the intersection regions of $k$ or more spheres is related to the $\mathrm{CDF}_{k\mathrm{NN}}(r)$ by
\begin{equation}
    \mathrm{CDF}_{k\mathrm{NN}}(r) = \frac{1}{V_0} \left|\bigcup_{\substack{J \subseteq \{1, 2, \dots, N_d\} \\ |J| \geq k}} \left( \bigcap_{i \in J} S_i \right)\right| 
    \label{eq:kNN_geometric}
\end{equation}
where $J$ is a set of data points containing at least $k$ of the $N_d$ total data points. The data points are represented by integers here, so the set of all data points is represented by $\{1, 2, \dots, N_d\}$. We denote the volume occupied by a set $S$ of points in space by $|S|$. Equation \ref{eq:kNN_geometric} gives the geometric interpretation of the $\mathrm{CDF}_{k\mathrm{NN}}(r)$ as the fraction of the total volume that is within the intersection of $k$ or more spheres of radius $r$ centered around the data points. Drawing spheres only around the data points, rather than at every point in space, greatly simplifies the interpretation of the $k$NN CDFs. This equivalence between the two methods provides a clearer and more flexible understanding of the  $k$NN CDFs. This will prove to be crucial as we shall see in section \ref{sec:knn_der} and \ref{sec:knn_mink}, allowing us to extract further geometric information about the dataset from the derivatives of the CDFs.
\newline\newline Let $V_k (r)$ be the volume within the intersection of \textit{exactly} $k$ spheres. These are the regions of a particular colour in figure \ref{fig:knn_geometric}. Then, equation \ref{eq:kNN_geometric} implies that 
\begin{equation}
    \mathrm{CDF}_{k\mathrm{NN}}(r) = \frac{1}{V_0}\sum_{i=k}^{\infty} V_i (r)
    \label{eq:at_least}
\end{equation}
\\
\noindent 
If we add all the $\mathrm{CDF}_{k\mathrm{NN}}$'s we will get a sum that includes the volume within only one sphere once, the volume within intersection of two spheres twice, the volume within intersections of three spheres thrice, and so on. Thus, appropriately summing both sides of equation \label{eq:knn_geometric} we get
\begin{equation}
    \sum_{k=1}^{\infty} \mathrm{CDF}_{k\mathrm{NN}}(r) = \frac{1}{V_0}\sum_{k=1}^{\infty} k V_k (r)
    \label{eq:sum_volumes}
\end{equation}
The volume within the intersection of  \textit{exactly} $k$ spheres can be given by subtracting the equation \ref{eq:sum_volumes} for $k+1$ from $k$ to obtain
\begin{equation}
\begin{aligned}
    V_k (r)  &= \left|\bigcup_{\substack{J \subseteq \{1, 2, \dots, N_d\} \\ |J| = k}} \left( \bigcap_{i \in J} S_i \right) \right| \\
    &= V_0\left\{ \mathrm{CDF}_{k\mathrm{NN}}(r) - \mathrm{CDF}_{(k+1)\mathrm{NN}}(r) \right\}
\end{aligned}
\label{eq:difference_of_cdfs}
\end{equation}
Using equations \ref{eq:CIC_def} and \ref{eq:difference_of_cdfs} we get the geometric interpretation of  Counts-in-Cells probabilities as 
\begin{equation}
    P_{k|V} = \frac{V_k (r) }{V_0}
    \label{eq:cic_knn}
\end{equation}
If we would have simply added the volume of all the spheres without taking into consideration their intersections, we would have exactly counted the regions of 2 sphere intersections twice, the regions of 3 sphere intersections thrice, and so on (using equation \ref{eq:sum_volumes}). This overcounting will allow us to exactly obtain the volume of all the spheres had they not intersected (see figure \ref{fig:knn_geometric}). Thus, 
\begin{equation}
    \sum_{k=1}^{\infty} \mathrm{CDF}_{k\mathrm{NN}}(r) =   \frac{4\pi r^3}{3} \frac{N_d}{V_0} = \frac{4\pi r^3}{3} \bar{n}
    \label{eq:sum_knn}
\end{equation}
where $\bar{n}$ is the mean number density of data points in our volume of interest. Observe that when $r$ is smaller than the smallest separation between two data points, we have non-intersecting spheres around data points. In this limit, the summation of CDFs in equation \ref{eq:sum_knn} reduces to just the $\mathrm{CDF}_{1\mathrm{NN}}$, and all other $\mathrm{CDF}_{k\mathrm{NN}} = 0$ for $k>1$. This observation aligns with our intuition that in the case of non-intersecting spheres, $V_0\times \mathrm{CDF}_{1\mathrm{NN}}$ is simply the sum of volumes of all the spheres around data points.
\\ \\
The $k$NN CDFs allow us to calculate the volume within spheres, intersections of exactly $k$ spheres and intersections of at least $k$ spheres around data points. In figure \ref{fig:knn_geometric}, the regions having a particular colour denote the points in space that are created by the intersection of \textit{exactly} $k$ spheres. The fraction of this volume gives us the CIC probability (equation \ref{eq:cic_knn}). Since the $k$NN CDFs give us the probability of finding \textit{at least} $k$ data points within a distance $r$, we can use equation \ref{eq:at_least} to interpret the 1NN CDF as the fraction of volume occupied by all the colours in figure \ref{fig:knn_geometric}, the 2NN CDF as the fraction of volume occupied by all colours except red, the 3NN CDF as the fraction of volume occupied by all colours except red and green, and so on. The 1NN region forms a union of spheres, the 2NN region forms a union of convex lenses (lenses formed by intersections of 2 spheres), and so on. Thus the higher $k$NN CDFs can be visually interpreted in terms of shapes created by intersections of multiple spheres.

\subsection{$k$NN cross-correlations and their geometric interpretation}
\label{sec:knn_cross}
\begin{figure*}
    \centering
    \begin{subfigure}[t]{0.48\linewidth}
        \centering
        \includegraphics[width=\linewidth]{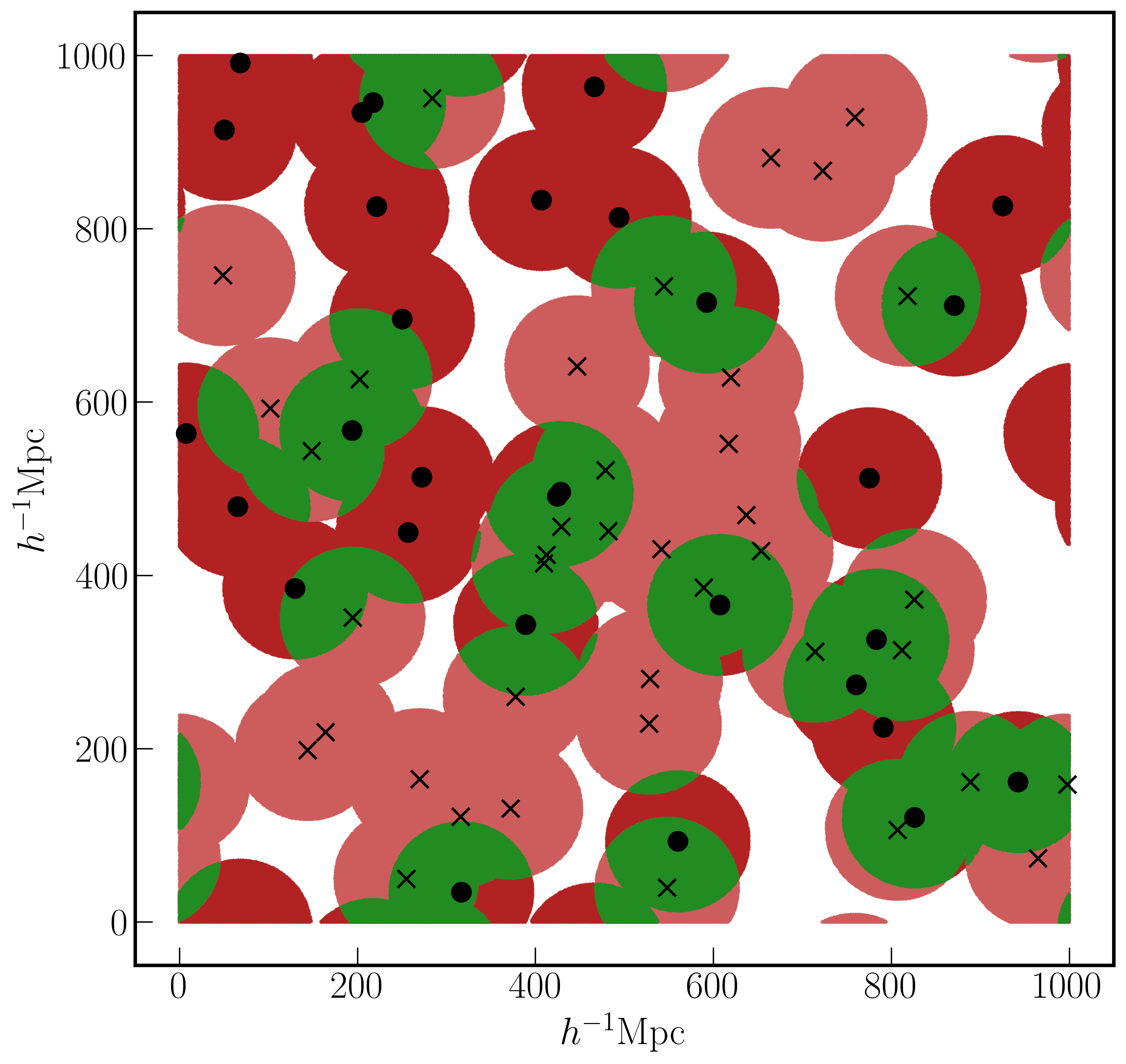}
        \caption{The figure illustrates the geometric interpretation for $\mathrm{CDF}_{1,1}(r)$. There are two different sets of data points here: circles and crosses. The area shaded in  dark (light) red  is the region where the 1NN distance from a circle (cross) data point is $\le r$. The area shaded in green is the region that is within a distance $\le r$ of at least one circle and one cross data point, corresponding to $\mathrm{CDF}_{1,1}(r)$.}
        \label{fig:cdf_11}
    \end{subfigure}
    \hfill
    \begin{subfigure}[t]{0.48\linewidth}
        \centering
        \includegraphics[width=\linewidth]{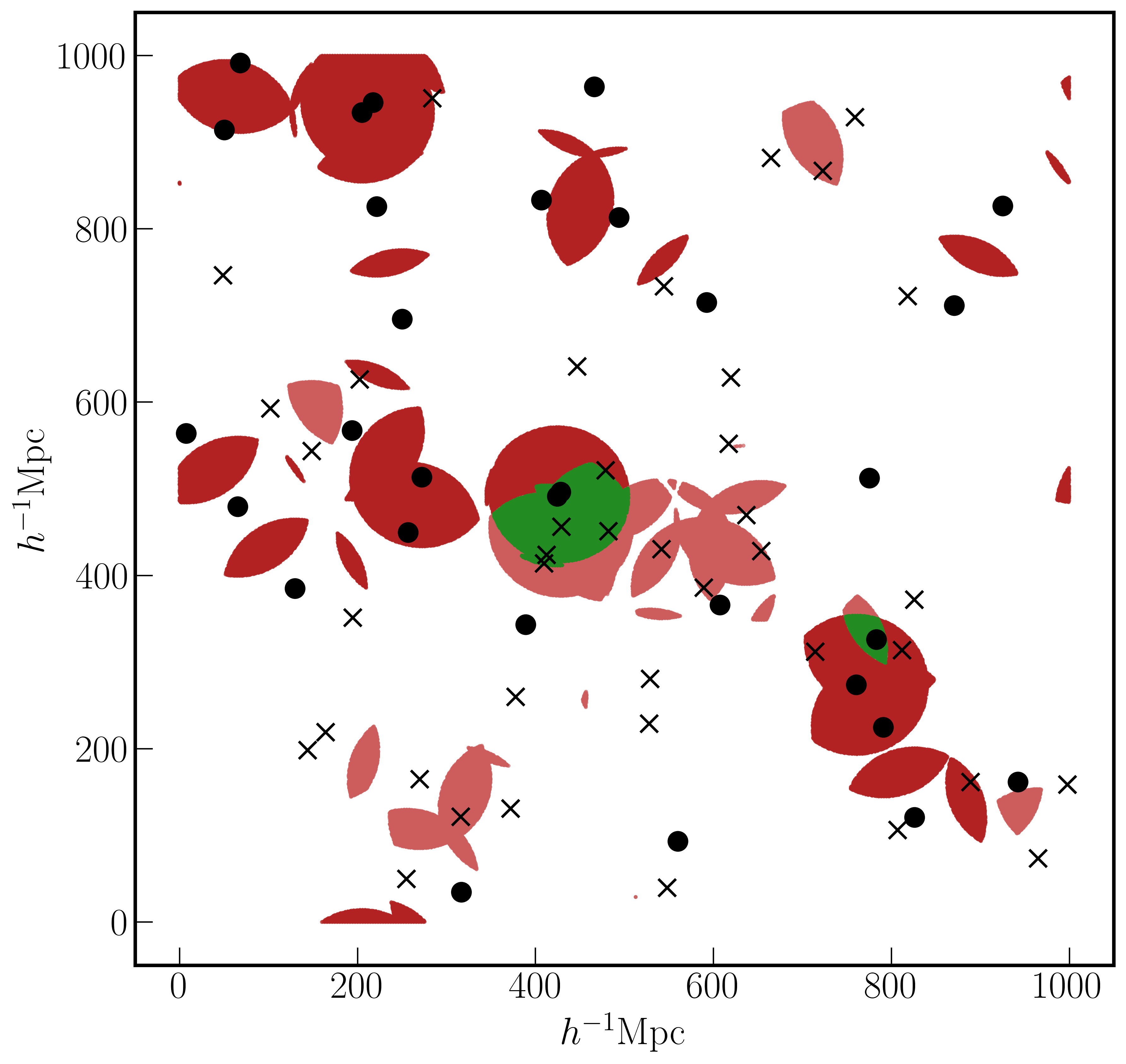}
        \caption{The figure illustrates the geometric interpretation for $\mathrm{CDF}_{2,3}(r)$. There are two different sets of data points here: circles and crosses. The area shaded in dark (light) red is the region where the 2NN (3NN) distance from a circle (cross) data point is $\le r$, representing the $\mathrm{CDF}_{2\mathrm{NN}}(r)$ ($\mathrm{CDF}_{3\mathrm{NN}}(r)$) for the corresponding tracer. The area shaded in green is the region that is within a distance $\le r$ of at least two circle and three cross data points, corresponding to $\mathrm{CDF}_{2,3}(r)$.}
        \label{fig:cdf_23}
    \end{subfigure}

    \caption{Geometric interpretation for the joint $\mathrm{CDF}_{k_1,k_2}(r)$.}
    \label{fig:crosscorr}
\end{figure*}

The $k$NN formalism can also be used to quantify the cross-clustering between two cosmological fields (\cite{10.1093/mnras/stab961}). Consider two sets of discrete tracers between which we want to measure the cross-clustering. Then, we can define their joint CDF using the $k$NN formalism as
\begin{equation}
    \mathrm{CDF}_{k_1,k_2}(r) = P(\ge k_1 \text{ of type 1}, \ge k_2 \text{ of type 2}\; \big| \; V=4\pi r^3/3)
\end{equation}
This gives the probability of finding at least $k_1$ data points of type 1 and at least $k_2$ data points of type 2 if we randomly draw spheres of radius $r$ in the space. The joint is equal to the product of the individual CDFs if the two fields are independent of each other. Thus, we can measure the true cross-clustering using 
\begin{equation}
    \psi^{(k_1,k_2)}(r) = \mathrm{CDF}_{k_1,k_2}(r) - \mathrm{CDF}^{(1)}_{k_1}(r)\cdot \mathrm{CDF}^{(2)}_{k_2}(r)
\end{equation}
where $\mathrm{CDF}^{(i)}_{k_i}(r)$ are the individual CDFs and the index on top denotes the field (type 1 or 2). Getting $\psi = 0$ implies that the two fields are independent, and a non-zero function $\psi(r)$ is a measure of the cross-clustering of the two fields. 
\newline\newline 
The joint $k$NN CDFs can be understood in terms of volumes of intersections of $k$NN regions for the two datasets. We start with the simplest case of $k_1=1, k_2=1$. The $\mathrm{CDF}_{1,1}(r)$ represents the fraction of points in the box that have at least one data point of each kind within a distance $r$. This is the region created by the intersections of the union of spheres around the two different kinds of data points (see figure \ref{fig:cdf_11}), as any point in these intersections is within a distance $r$ of at least one data point of each kind. We can extend this to higher $k$'s, but instead of the union of spheres we will use the union of lenses ($k=2$) or other shapes corresponding to the regions from where the appropriate $k$NN distances are $<r$ (see figure \ref{fig:cdf_23}). 

\subsection{Derivatives of $k$NN CDFs}
\label{sec:knn_der}
\begin{figure}
    \centering
    \includegraphics[width=0.5\linewidth]{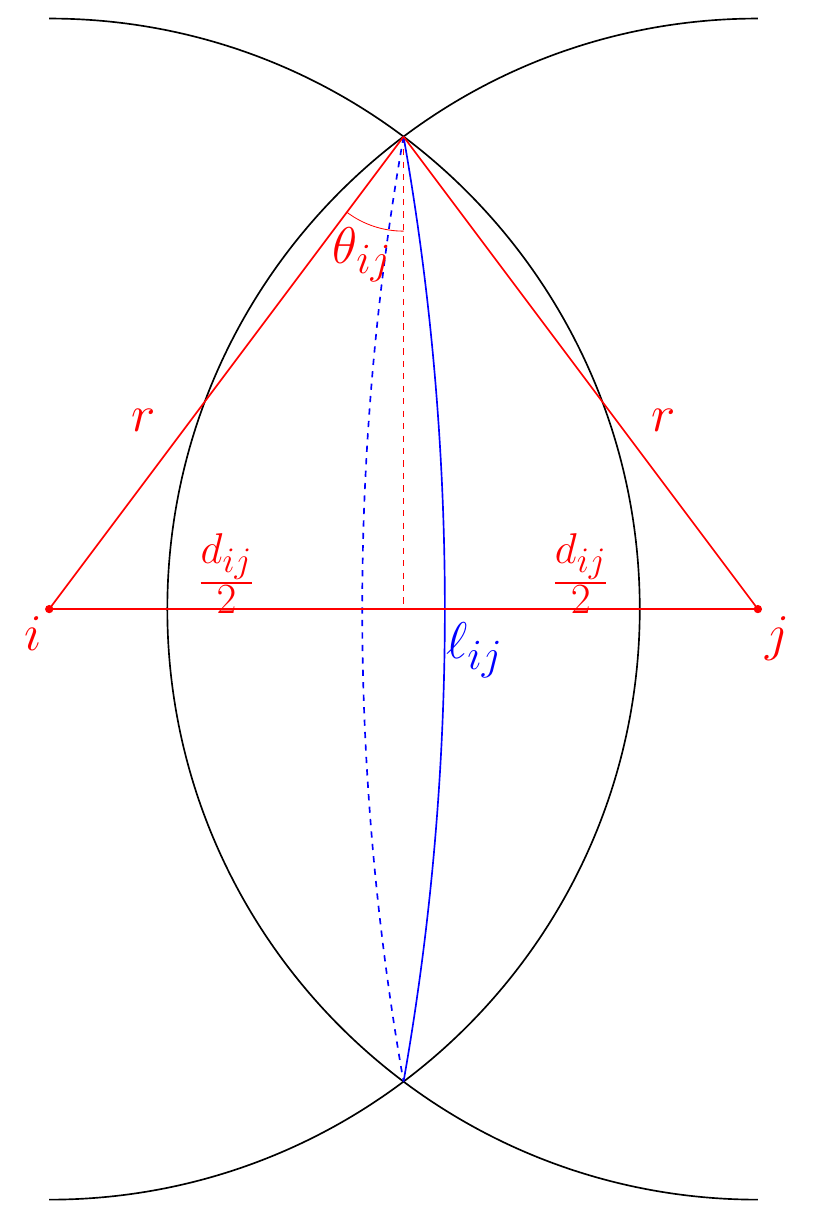}
    \caption{This figure shows the intersection circle created by the the intersection of spheres around two points $i$ and $j$. The length of the circle is $\ell_{ij}$. But for intersection of three or more spheres, the circle is not fully on the surface, and we only need to consider the length of the arc that is on the surface.}
    \label{fig:int_spheres}
\end{figure}

Having established a geometric interpretation for the $k$NN CDF at a particular radius $r$, we can also ask if the rate of change of the value of the CDF at any radius $r$ can also be interpreted in some analogous manner. The derivatives of the $k$NN CDFs also contain geometric information that is very useful for measuring clustering in a cosmological dataset. 
The area $A(r)$ of the union  of spheres around data points is proportional to the first derivative of the $\mathrm{CDF}_{1\mathrm{NN}}$ (see Appendix \ref{sec:arc}):
\begin{equation}
    A(r) = V_0\times \frac{d}{dr}\mathrm{CDF}_{1\mathrm{NN}}(r)
\end{equation}
This is intuitive, as with a small increase in radius $\Delta r$, the volume of the union of spheres should change by $A(r) \Delta r$ upto linear order. There are corrections to this due to the sharp edges created at the intersections of spheres where the surface is not smooth. But we have shown that these corrections only occur at higher-orders of $\Delta r$, a proof is in appendix \ref{sec:arc}. For a collection of non-intersecting spheres of radius $r$, knowing the total volume gives us the total number of spheres, and there is no other information we can extract from such a setup. But when the spheres intersect, the total volume and surface area of the union of spheres depend on the geometric arrangement of the sphere centers in space. Thus, knowing both the volume and area from the CDFs is useful for performing a clustering analysis in a cosmological context. Further information can be obtained if we can know about the angles subtended at the intersections of spheres by the centers, or the lengths of the arcs/circles created at the intersections of spheres. 
\\ \\
The second derivative of the $\mathrm{CDF}_{1\mathrm{NN}}$ reveals information about the intersections of spheres through angles and arc lengths (see Appendix \ref{sec:arc}):
\begin{equation}
    V_0\times\frac{d^2\mathrm{CDF}_{1\mathrm{NN}}(r)}{dr^2} = 2 \left[\frac{A(r)}{r} - \frac{1}{2}\sum_{i,j}\ell_{ij}\tan{\theta_{ij}}\right]
    \label{eq:der2_arclength}
\end{equation}
where the summation is over all pairs of data points $i,j$, spheres around which intersect along circular arcs of arc length $\ell_{ij}$, and the half angle subtended by the two points at the intersection arc is $\theta_{ij}$ (see figure \ref{fig:int_spheres}). Note that $\ell_{ij}$ is the length of the arc that is on the surface of the spheres, and for any circle created at the intersection of spheres only the length of the arc on the surface is considered in the summation. Since the summation goes over all pairs, it double counts. Hence, there is an additional factor of $\frac{1}{2}$ to correct the summation. This gives us information about the angles and arc lengths created at the intersections of two spheres at a time. Combining this with the total volume and surface area of the union of spheres can help in studying the geometric arrangement of the data points in space. 
\\ \\
Further information on the solid angles created at the intersection points of three spheres will be useful for performing a clustering analysis. Since the intersections of four or more spheres at a single point is an extremely rare event, we stop at the intersection points of three spheres. It is in this light that we state the correspondence between the Euler characteristic and $\frac{d^3V(r)}{dr^3}$. The reasoning behind this method is explained in Appendix \ref{sec:v_eps}. The Euler characteristic is a topological invariant, stable under small perturbations and deformations in the shape. For a single solid sphere, the Euler characteristic $\chi = 1$. For a union of multiple spheres, we can compute $\chi$ using the additivity property (equation \ref{eq:add}. It can also be written as the alternating sum of Betti numbers $\beta_{i}$. Thus, in 3 dimensions, we have
\begin{equation}
    \chi = \beta_0 - \beta_1 + \beta_2
\end{equation}
where $\beta_0$ counts the number of connected components, $\beta_1$ the number of holes through which we can pass through, and $\beta_2$ the number of voids or closed holes (like within a spherical shell). 
\\ \\
Since intersections of more than three spheres at one point is an extremely rare event, we can ignore their contribution to $\chi$. Then, it can be written in terms of the arc lengths, angles at intersection of two spheres and solid angles at intersection of three spheres (\cite{1994A&A...288..697M}) as 
\begin{equation}
\begin{aligned}
    \chi &= \frac{1}{4\pi} \left[ \frac{A(r)}{r^2} - \frac{1}{2} \sum_{i,j} \frac{d_{ij} \ell_{ij}}{r \sqrt{r^2 - (d_{ij}/2)^2}} + \frac{1}{6}\sum_{i,j,k} \Omega_{ijk}\right] \\
    &= \frac{1}{4\pi} \left[ \frac{A(r)}{r^2} -  \sum_{i,j} \frac{\ell_{ij} \tan \theta_{ij} }{r} + \frac{1}{6}\sum_{i,j,k} \Omega_{ijk}\right]
\end{aligned}
    \label{eq:chi_arclength}
\end{equation}
where $\Omega_{ijk}$ is the solid angle created at the intersection point of three spheres with centers $i,j,k$. The extra factors of $1/2$ and $1/6$ are to correct for overcounting performed in the summation. 
\\ \\
\noindent 
We find that the third derivative of the 1NN CDF is closely linked to the Euler characteristic. Since the Euler characteristic is a discretely varying quantity, it changes discontinuously whenever an intersection happens, a hole is created or closed, or the topology of the union of spheres changes. It only takes integral values, whereas $\frac{d^3V(r)}{dr^3}$ does not take only integral values. But it turns out (see appendix \ref{sec:v_eps}) that the $\frac{d^3V(r)}{dr^3}$ also changes discontinuously at intersections, when a hole is created or closed, or the topology of the union of spheres changes. In fact, between every such change, where $\chi$ remains a constant, $\frac{d^3V(r)}{dr^3}$ asymptotes to $8\pi\chi$ in that region (equation \ref{eq:asymptote}). Since $\frac{d^3V(r)}{dr^3}$ is a smooth and infinitely differentiable function between every change in topology, we can fit a Laurent series in each such region (equation \ref{eq:laurent}) to extract the constant value to which $\frac{d^3V(r)}{dr^3}$ is asymptoting towards, hence obtaining $\chi$ in that region. Thus, we claim that $\frac{d^3V(r)}{dr^3}$ and $\chi$ have similar information content, just packed in a different way. Given infinite precision, we can compute $\chi$ from $\frac{d^3V(r)}{dr^3}$. A pseudocode for the same has been presented in algorithm \ref{alg:euler}.
\begin{algorithm}
\caption{Algorithm to compute array containing $\chi$ values, using $\frac{d^3 V(r)}{dr^3}$ as a function of $r$ between some $r_{\text{min}}$ and $r_{\text{max}}$, given the volume within spheres around data points $V(r)$. This algorithm assumes an arbitrarily precise knowledge of $V(r)$, so that we can check for the continuity of its derivative. The final array $\text{euler}$ is such that $\text{euler}[i]$ contains the value of $\chi$ between $\text{disc}[i]$ and $\text{disc}[i+1]$ (or $r_{\text{max}}$ for the last value).}
\label{alg:euler_array}
\begin{algorithmic}[1]
\State \textbf{Input:} $r_{\text{min}}$, $r_{\text{max}}$, $V(r)$ (predefined as a function of $r$)
\State \textbf{Output:} $\text{euler}$ (array), $\text{div}$ (array)
\State Initialize $\text{disc} \gets \{r_{\text{min}}\}$ \Comment{Array to store discontinuities}
\State Initialize $\text{euler} \gets \{\}$ \Comment{Array to store Euler terms}
\State $r \gets$ arbitrarily dense array from $r_{\text{min}}$ to $r_{\text{max}}$
\For{each $x \in r$}
    \If{$\left(\frac{d^3 V(r)}{dr^3}\Big|_{r\rightarrow x^-} \neq \frac{d^3 V(r)}{dr^3}\Big|_{r\rightarrow x^+}\right)$}
        \State Append $x$ to $\text{disc}$
    \EndIf
\EndFor
\State $N \gets \text{len}(\text{disc})$
\For{$i = 0$ to $N-1$}
    \State $x_1 \gets \text{disc}[i]$
    \If{$i < N-1$}
        \State $x_2 \gets \text{disc}[i+1]$
    \Else
        \State $x_2 \gets r_{\text{max}}$
    \EndIf
    \State Fit $\frac{d^3 V(r)}{dr^3}$ in the range $x_1 < r < x_2$ using the model:
    \[
    f(r) = c_0 + \frac{c_1}{(r - x_1)} + \frac{c_2}{(r - x_1)^2} + \frac{c_3}{(r - x_1)^3} + \dots
    \]
    by minimizing the difference between $\frac{d^3 V(r)}{dr^3}$ and $f(r)$ to obtain the best-fit coefficients $c_0, c_1, c_2, \dots$
    \State Append $\frac{c_0}{8\pi}$ to $\text{euler}$
\EndFor
\end{algorithmic}
\label{alg:euler}
\end{algorithm}
\\ \\
The Euler characteristic as a function of $r$ can distinguish between datasets that have different kinds of structures. For example, a cosmology that shows a higher fraction of mass in linear structures like filaments will behave differently to a cosmology which has more voids. These topological differences are well captured by the Euler characteristic. Since we have shown that the third derivative of the 1NN CDF contains similar information, the 1NN CDF and its derivatives should also be able to differentiate between these different topological structures in cosmological datasets. 
\\ \\
The derivatives of the 1NN CDF provide valuable information about the geometry of the data points through spheres and their intersections. A similar approach can be useful for higher $k$'s, which might lead to additional information that helps constrain the cosmological parameters even better. In fact, the $\mathrm{CDF}_{k\mathrm{NN}}(r)$ is readily interpretable as the volume within the region where the distance to the $k^{\mathrm{th}}$ nearest neighbour data point is at most $r$, and the first derivative of the CDF is interpretable as the area of this region. Physically, these regions correspond to the intersection of lenses and other shapes, depending on $k$. Figure \ref{fig:cdf_23} shows how the 2NN and 3NN regions for a set of data points might look like. We have used the derivatives of $k$NN CDFs (along with the CDFs themselves) for $k=1,2,3,4$ to constrain cosmological parameters, and the results are presented in section \ref{sec:fisher}. It is important to note that while the extra information about the area, angles and topology of the union of spheres is present in the derivatives of the CDF (and thus theoretically present in the full shape of the CDF itself), using the derivatives separately as data vectors makes the geometric information more accessible for a realistic clustering analysis.
\\ \\
In sections \ref{sec:knn_geo}, \ref{sec:knn_cross} and \ref{sec:knn_der} we have explored the geometric interpretations of the $k$NN CDFs, the cross-correlation joint CDFs and the derivatives of the $k$NN CDFs. Table \ref{tab:summary_knngeo} summarizes these geometric interpretations.
\begin{table*}
    \centering
    \caption{Geometric Interpretations of the $k$NN CDFs and various quantities related to them}
    \begin{tabular}{|l|l|}
    \hline
    \textbf{Quantity} ($\times V_0$) & \textbf{Geometric Interpretation}\\
    \hline\hline
    $\mathrm{CDF}_{k\mathrm{NN}}(r)$ & Volume of region within which $\ge k$ data points are located within a distance $r$\\
    \hline
    $\mathrm{CDF}_{1\mathrm{NN}}(r)$ & Volume of union of spheres of radius $r$ around data points\\
    \hline 
    $\mathrm{CDF}_{k_1,k_2}(r)$ & Volume of region within which $\ge k_1, k_2$ data points of type 1, 2 are located within a distance $r$\\
    \hline
    $\mathrm{CDF}_{1,1}(r)$ & Volume of intersections of spheres of radius $r$ around data points of type 1 and type 2, \\ & forming a union of  convex lens shaped regions\\
    \hline
    $\frac{d}{dr}\mathrm{CDF}_{k\mathrm{NN}}(r)$ & Surface Area of region within which $\ge k$ data points are located within a distance $r$\\
    \hline
    $\frac{d}{dr}\mathrm{CDF}_{1\mathrm{NN}}(r)$ & Surface Area of union of spheres of radius $r$ around data points\\
    \hline
    $\frac{d^2}{dr^2}\mathrm{CDF}_{1\mathrm{NN}}(r)$ & Contains information about the angles and arcs created by the intersections of spheres of radius $r$ around data points\\
    \hline 
    $\frac{d^3}{dr^3}\mathrm{CDF}_{1\mathrm{NN}}(r)$ & Contains information about the solid angles created by the intersections of three spheres of radius $r$ around data points, \\ & also contains information about the Euler Characteristic of the union of spheres   \\ \hline \hline
\end{tabular}
    \label{tab:summary_knngeo}
\end{table*}

\section{1NN CDF and Minkowski Functionals}
\label{sec:knn_mink}
In sections \ref{sec:knn_geo} and \ref{sec:knn_der}, we constructed spheres of radius $r$ around data points to compute $\mathrm{CDF}_{1\mathrm{NN}}(r)$ and its derivatives. This approach of measuring the clustering of cosmological tracers by drawing spheres around them is reminiscent of the Germ-Grain Minkowski Functionals (\cite{1994A&A...288..697M}). These are geometric quantities associated with the union of spheres around the data points, and the function $V(r)$ is the same as the first Minkowski Functional $W_0 (r)$. The other Minkowski Functionals measure geometric and topological quantities like the surface area, integrated mean curvature and Euler characteristic of this set of spheres around data points. In this section, we establish a correspondence between these four Minkowski Functionals and the $\mathrm{CDF}_{1\mathrm{NN}}(r)$ and its three derivatives.
\\ \\
 Minkowski Functionals have been used on cosmological simulations (e.g., \cite{Schmalzing:1995qn}) and observational data sets (e.g., \cite{10.1093/mnras/stu1118}, \cite{2025arXiv250101698L}) to characterize the clustering of the large scale structure of the universe, giving better constrains than the two-point statistics. Additionally, Minkowski Functionals and Tensors have been used to study continuous cosmological fields like the CMB (e.g., \cite{2013PhRvD..88d1302F}, \cite{2017PhLB..771...67C}), galaxy density field (e.g., \cite{2023JCAP...09..037L}, \cite{2024PhRvD.109h3537J}) and the 21cm brightness temperature field (e.g., \cite{2019JCAP...09..053K}, \cite{2021JCAP...05..026K}).
\newline\newline 
A mathematical definition of the Minkowski Functionals, along with how to compute them for spheres around data points, is explained in appendix \ref{sec:v_eps}. The primary definition of the Minkowski Functionals is for compact convex objects. In three dimensions, there exist four Minkowski Functionals. These four Minkowski Functionals are closely related to some geometric and topological properties of the compact, convex object $\mathcal{K}$ we are working with:
\begin{equation}
\begin{aligned}
&W_0(\mathcal{K}) = V(\mathcal{K}); 
\hspace{2cm} \mathrm{V:Volume} \\
&3W_1(\mathcal{K}) = S(\mathcal{K});
\hspace{1.9cm} \mathrm{S:Surface\hspace{0.1cm} Area}\\
&3W_2(\mathcal{K}) = H(\mathcal{K});
\hspace{1.8cm} \mathrm{H:Integral\hspace{0.1cm}Mean\hspace{0.1cm}Curvature}\\
&3W_3(\mathcal{K}) = 4\pi\chi(\mathcal{K});
\hspace{1.5cm} \mathrm{\chi:Euler\hspace{0.1cm} Characteristic}\\
\end{aligned}
\label{eq:minkowski_geometric}
\end{equation}
 The integral mean curvature is defined as 
\begin{equation}
    H(\mathcal{K}) = \frac{1}{2}\int_{\partial \mathcal{K}} \left(\frac{1}{R_1}+\frac{1}{R_2}\right)ds
\end{equation}
where $R_1$ and $R_2$ are the principal radii of curvature (the maximal and minimal) and $\partial \mathcal{K}$ is the surface of $\mathcal{K}$. The Euler characteristic is a topological invariant, and for a compact convex object it is simply 1. Thus, $\chi(\mathcal{K}) = 1$. 
\\ \\
The Germ-Grain Minkowski Functionals are simply the Minkowski Functionals evaluated on a set of growing spheres around the \textit{data points}, as a function of the sphere radius $r$. In order to compute them, we need to extend the definition of the Minkowski Functionals to non-convex objects. This can be done by fragmenting our non-convex object as a union of convex parts, and using the additivity property on them to compute the Minkowski Functionals of the entire object (see equation \ref{eq:add}). In this section, we present the results of these calculations directly, and explain how these calculations can be done in appendix \ref{sec:v_eps}.
\\ \\
The first Minkowski Functional $W_0(r)$ is simply the volume within the union of spheres of radius $r$ drawn around all the data points (equation~\ref{eq:minkowski_geometric}). So we can rewrite equation~\ref{eq:W0_1} to get our first Minkowski-Nearest Neighbour relation:
\begin{equation}
    W_0(r) = V_0\times \mathrm{CDF}_{1\mathrm{NN}}(r)
    \label{eq:W0_2}
\end{equation}
The derivations relating $W_1$, $W_2$ and $W_3$ to $\frac{d}{dr}\mathrm{CDF_{1\mathrm{NN}}}$, $\frac{d^2}{dr^2}\mathrm{CDF_{1\mathrm{NN}}}$ and $\frac{d^3}{dr^3}\mathrm{CDF_{1\mathrm{NN}}}$ has been presented in Appendix \ref{sec:arc} and \ref{sec:v_eps}. Here, we state the main results and their implications for analysing cosmological data.
\\ \\
The total surface area of the union of spheres $A(r)$, is related to both $W_1(r)$ and $\frac{d}{dr} \mathrm{CDF}_{1\mathrm{NN}}(r)$:
\begin{equation}
\begin{aligned}
     &A(r) = V_0\times \frac{d}{dr}\mathrm{CDF}_{1\mathrm{NN}}(r) \; ; \\
     &A(r) = 3 W_1 (r) \; ; \\
    \Rightarrow & W_1 (r) = \frac{V_0}{3}\times \frac{d}{dr}\mathrm{CDF}_{1\mathrm{NN}}(r)
\end{aligned}  
\end{equation}
Thus, $\mathrm{CDF}_{1\mathrm{NN}}$ and its first derivative can exactly give us the first two Minkowski Functionals. For $W_2, W_3$, the correspondence is not exact. But as we will show, both the Minkowski Functionals and derivatives of $\mathrm{CDF}_{1\mathrm{NN}}$ contain essentially the same information. 
\\ \\
The next Minkowski Functional, which is proportional to the integrated mean curvature, is sensitive to both the total surface area and the lengths of the arcs created at the intersection of two spheres through the following relation (see \cite{1994A&A...288..697M}):
\begin{equation}
    W_2 (r) = \frac{1}{3} \left[\frac{A(r)}{r} - \frac{1}{2}\sum_{i,j}\ell_{ij}\theta_{ij}\right]
    \label{eq:mean_curv_arclength}
\end{equation}
where the summation is over all pairs of points $i,j$ whose spheres intersect along circular arcs of arc length $\ell_{ij}$, and the half angle subtended by the two points at the intersection arc is $\theta_{ij}$. Since the summation goes over all pairs, it double counts. Hence, there is an additional factor of $\frac{1}{2}$ to correct the summation. Since $V(r)$ and $\mathrm{CDF}_{1\mathrm{NN}}$ are related by a constant factor (equation \ref{eq:W0_1}), we choose to work with $V(r)$ henceforth. Rewriting equation \ref{eq:der2_arclength} in terms of $V(r)$ we obtain:
\begin{equation}
    \frac{d^2V(r)}{dr^2} = 2 \left[\frac{A(r)}{r} - \frac{1}{2}\sum_{i,j}\ell_{ij}\tan{\theta_{ij}}\right]
    \label{eq:der2_arclength_vol}
\end{equation}
This equation looks almost like equation~\ref{eq:mean_curv_arclength}, only differing in $\tan{\theta_{ij}}$ instead of $\theta_{ij}$. We can alternatively write equations~\ref{eq:mean_curv_arclength} and \ref{eq:der2_arclength_vol} in terms of $f_{ij}$'s instead of $\ell_{ij}$'s:
\begin{equation}
    \begin{aligned}
       & W_2 (r) = \frac{1}{3} \left[\frac{A(r)}{r} - \frac{1}{2}\sum_{i,j} 2\pi r f_{ij}  \theta_{ij} \cos{\theta_{ij}}\right] \\
       & \frac{d^2V(r)}{dr^2} = 2 \left[\frac{A(r)}{r} - \frac{1}{2}\sum_{i,j} 2\pi r f_{ij}  \sin{\theta_{ij}}\right]
    \end{aligned}
\end{equation}
where $f_{ij}$ is the fraction of the complete circle that appears on the surface. We claim that both $W_2(r)$ and $\frac{d^2V(r)}{dr^2}$ contain similar information. This is because
\begin{itemize}
    \item both functions are dependent on $A(r)$, $\ell_{ij}$ and $\theta_{ij}$
    \item both functions only differ in the $\theta_{ij}$ dependence
    \item both functions asymptote to the same value for large radius ($\theta_{ij}$ decreases as $r$ increases)
    \item both summations only add new terms when a new circle/arc appears on the surface, created by the intersection of spheres
\end{itemize}
We further compare the cosmological constraining power of the two functions in section \ref{sec:fisher}.
\\ \\
The last Minkowski Functional $W_3(r)$ is proportional to the Euler Characteristic $\chi$. We have already shown a correspondence between $\frac{d^3 V(r)}{dr^3}$ and $\chi$ in section \ref{sec:knn_der}. We also presented an algorithm (see algorithm \ref{alg:euler_array}) to compute $\chi$ from $\frac{d^3 V(r)}{dr^3}$ if we have infinite precision. But in realistic data where we only take finitely many bins, this is not possible. Hence, for realistic cosmological analyses (like the Fisher analysis we have done in section \ref{sec:fisher}), $\chi$ might still be a more constraining statistic. However, it is computationally very expensive to compute, compared to the $k$NN CDFs. And as we shall show in section \ref{sec:fisher}, the 1NN CDF and its derivatives provide roughly similar cosmological parameter constraints as the  Minkowski Functionals. Even the directions of the elipses in the contour plots (figure \ref{fig:1nnder_vs_mink}) are similar, reinforcing our claims that the derivatives of the 1NN CDF and the Minkowski Functionals are closely linked. 
\\ \\
It is to be noted that the $k$NN CDFs and Minkowski Functionals are defined in completely different ways, one from counting data points within a distance $r$ from random points in space, another from drawing spheres of radius $r$ around data points and measuring certain geometric quantities. In the way they are originally defined, the $k$NN CDFs capture the probabilities associated with the spatial arrangement of data points, whereas the Minkowski Functionals capture the geometry of this spatial arrangement. Thus, the fact that they end up with very similar geometric interpretations and comparable information is quite remarkable. 
\\ \\
Although we have shown a correspondence between the 1NN CDF and the Minkowski Functionals (including $\chi$), incorporating higher $k$NN CDFs does not allow us to cleanly express the Betti numbers. With advanced data analysis techniques like Persistent Homology beginning to be used to analyze cosmological datasets, we believe that using the 3D map of query points' $k$NN distances from the data points will allow us to get constraints competitive to the topological data analysis techniques. This is because the $k$NN CDF is a compressed version of this 3D map, where we eliminate all the spatial information. Thus, if we use the more informative 3D maps which has the spatial information, we believe we can compete with the topological data analysis techniques like Betti Numbers and Persistent Homology. This methodology is under investigation by \cite{ChatterjeeInPrep}, and a future paper will explore the applications of machine learning techniques to extract information from these 3D $k$NN maps.

\section{Fisher analysis}
\label{sec:fisher}

\begin{figure*}
    \centering
    \includegraphics[width=\textwidth]{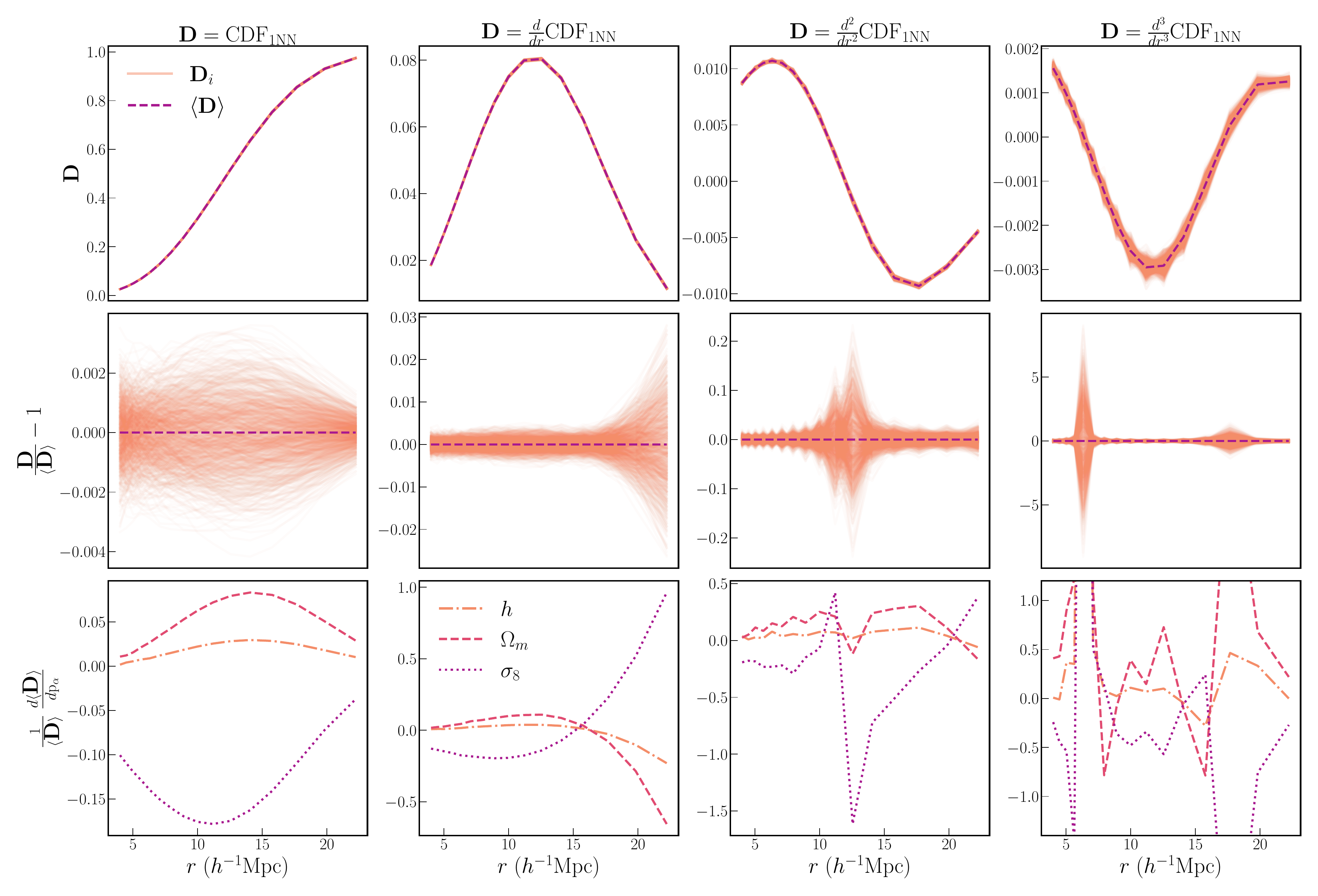}  
    \caption{The top panel is for the data vector of 1NN CDF and its three derivatives. The four parts are plotted in the four columns, but the Fisher analysis has been done by combining all these into a single data vector. In the first two rows, the light orange curves represent the data vectors from 1000 fiducial realizations of the \textsc{Quijote} simulations, and the dashed purple curve represents the mean of these 1000 fiducial data vectors. The bottom pannel shows the derivatives of the data vectors with respect to the cosmological parameters, normalized by the mean of the 1000 fiducial data vectors.}
    \label{fig:data_vectors}
\end{figure*}

In this section, we use the Fisher matrix formalism to compare the constraining power of the $k$NN CDFs, their derivatives, and the Minkowski Functionals. Assuming that the different realizations of our data vector are Gaussian distributed around the mean, the Fisher matrix formalism allows us to compute the variance and covariance of these Gaussian distributions and hence estimate the constraints that the various statistics (and their corresponding data vectors) can achieve. This approach is particularly useful for statistics with error bars that are difficult to compute analytically. To estimate these uncertainties, we compute the data vectors using a large number of realizations of cosmological simulations. It is important to note that while this method provides an estimate of the covariance, it does not yield the mean values of the parameters. In our analysis, we therefore assume that the mean values correspond to the fiducial parameters.
\\ \\
The elements of the Fisher matrix $\textbf{F}$ are given by
\begin{equation}
    \textbf{F}_{\alpha\beta} = \sum_{i,j} \frac{\partial \textbf{D}_i}{\partial p_\alpha} \left( \textbf{C}^{-1}\right)_{ij} \frac{\partial \textbf{D}_j}{\partial p_\beta}
    \label{eq:fisher}
\end{equation}
where $\textbf{D}_i$ are the elements of the data vector, $p_\alpha$ are the cosmological parameters we are trying to constrain and $\textbf{C}$ is the covariance matrix for the data vector at the fiducial values of the parameters. Note that this is different from the covariance matrix for the cosmological parameters, which is estimated by $\textbf{F}^{-1}$ under the Gaussian assumption we made earlier.
\\ \\
The $1\sigma$ constraint for the parameter $p_\alpha$ can be obtained by inverting the Fisher matrix:
\begin{equation}
    \sigma_\alpha = \sqrt{\left(\textbf{F}^{-1}\right)_{\alpha\alpha}}
\end{equation}
The raw covariance matrix has elements 
\begin{equation}
    \tilde{\textbf{C}}_{ij} = \langle \left(\textbf{D}_i - 
\langle \textbf{D}_i \rangle \right)\left(\textbf{D}_j - 
\langle \textbf{D}_j \rangle \right)\rangle
\end{equation}
where $\langle ... \rangle$ denotes an average over all realization. To compute the Fisher matrix however, we need to modify this raw covariance matrix to make it an unbiased estimator of the true covariance matrix. The correct inverse covariance matrix to be used in equation~\ref{eq:fisher} has an additional Hartlap factor (\cite{2007A&A...464..399H}): 
\begin{equation}
    \textbf{C}^{-1} = \left(\frac{n-p-2}{n-1} \right)\tilde{\textbf{C}}^{-1}
    \label{eq:hartlap}
\end{equation}
where $n$ is the total number of realizations used to compute the covariance matrix, and $p$ is the number of elements in the data vector. Multiplying this Hartlap factor ensures that $\textbf{C}$ (and not $\tilde{\textbf{C}}$) is an unbiased estimator for the true covariance matrix.
\newline\newline
We use the \textsc{Quijote}\footnote{\url{https://quijote-simulations.readthedocs.io/en/latest/}} simulations (\cite{2020ApJS..250....2V}) to compute the covariance matrices and the derivatives of the data vectors. The \textsc{Quijote} simulations are $N$-body cosmological simulations performed in boxes of volume $(1 h^{-1} \mathrm{Gpc})^3$. The simulations we used were CDM only simulations, with $512^3$ CDM particles. The simulations allow easy computation of the derivative around a fiducial cosmology, since there are simulations with parameter values marginally above and below the fiducial values, only changing one parameter at a time (see table~\ref{tab:simulations}). The cosmological parameters we use for this analysis are $\{ h,\Omega_m,\sigma_8 \}$. We use simulation snapshots at $z=0$ for our analysis. 
\\ \\
The aim of our Fisher analysis was to validate the theoretical correspondence between the information content in the $k$NN CDFs and derivatives, and the Minkowski Functionals. For this, we needed to use a data set in which the systematic errors are minimal and the results converge to the theoretical expectations.
\\ \\
For Fisher analyses using dark matter halo catalogs of Quijote simulations, the derivatives of the data vectors with respect to the cosmological parameters have been shown to be biased (\cite{Coulton_2023}), not converging even with 1000 realizations (500 for positive and 500 for negative perturbations). Thus, we opted to use CDM particles rather than dark matter halos in our analysis as the derivatives and covariance matrices converge with fewer realizations. We have checked the convergence explicitly by varying the number of simulations used to calculate the covariance matrices. The matrices (scaled between $-1$ and $+1$) for a few of the data vectors are shown in appendix \ref{sec:cov} (figure \ref{fig:correlation}).
\begin{table}
    \centering
    \caption{The different simulations used to compute the data vectors and their derivatives.}
    \begin{tabular}{|l|c c c}
    \hline
    Name & $h$ & $\Omega_m$ & $\sigma_8$ \\
    \hline
    fiducial & 0.6711 & 0.3175 & 0.834\\
    h\_p & 0.6911 & 0.3175 & 0.834\\
    h\_m & 0.6511 & 0.3175 & 0.834\\
    Om\_p & 0.6711 & 0.3275 & 0.834\\
    Om\_m & 0.6711 & 0.3075 & 0.834\\
    s8\_p & 0.6711 & 0.3175 & 0.849\\
    s8\_m & 0.6711 & 0.3175 & 0.819\\
    \hline
\end{tabular}
    \label{tab:simulations}
\end{table}
\\ \\
The fiducial data vectors (for computing the covariance matrix $\textbf{C}$) were computed for 1000 fiducial realizations. To compute the derivative of the data vector (say, with respect to $h$) we computed the mean of the data vectors for 100 realizations for both the simulations h\_p and h\_m, and divided their difference by the change in $h$. For computing all the data vectors, we randomly down-sampled $10^5$ DM particles from the total $512^3$ DM particles in the box, giving us a mean number density of 
\begin{equation}
    \bar{n} = \frac{10^5\; \textrm{particles}}{(1000 \;h^{-1}\mathrm{Mpc})^3} = 10^{-4} \; \frac{\textrm{particles}}{(h^{-1}\mathrm{Mpc})^3}
\end{equation}
Note that both the $k$NN CDFs and Minkowski Functionals are highly sensitive to the mean number density of tracers. It is important to choose $\bar{n}$ appropriately to so that the CDFs and Minkowski Functionals probe the range of length scales one is interested in. 
\\ \\
\noindent We compute the following data vectors:
\begin{itemize}
    \item $\mathrm{CDF}_{1\mathrm{NN}}$ and its first three derivatives with respect to $r$, each at 16 bins (64 dimensional vector)
    \item $W_0, W_1, W_2, W_3$ each at 16 bins (64 dimensional vector)
    \item $\mathrm{CDF}_{1\mathrm{NN}},\mathrm{CDF}_{2\mathrm{NN}},\mathrm{CDF}_{3\mathrm{NN}},\mathrm{CDF}_{4\mathrm{NN}}$, each at 16 bins (64 dimensional vector)
    \item $\mathrm{CDF}_{1\mathrm{NN}},\mathrm{CDF}_{2\mathrm{NN}},\mathrm{CDF}_{3\mathrm{NN}},\mathrm{CDF}_{4\mathrm{NN}}$ and their first derivatives with respect to $r$, each at 16 bins (128 dimensional vector)
    \item Parts of these data vectors, treated as individual data vectors of smaller dimension (for example, just the $\mathrm{CDF}_{1\mathrm{NN}}$ in 16 bins)
\end{itemize}
\begin{figure}
    \centering
    \includegraphics[width=\linewidth]{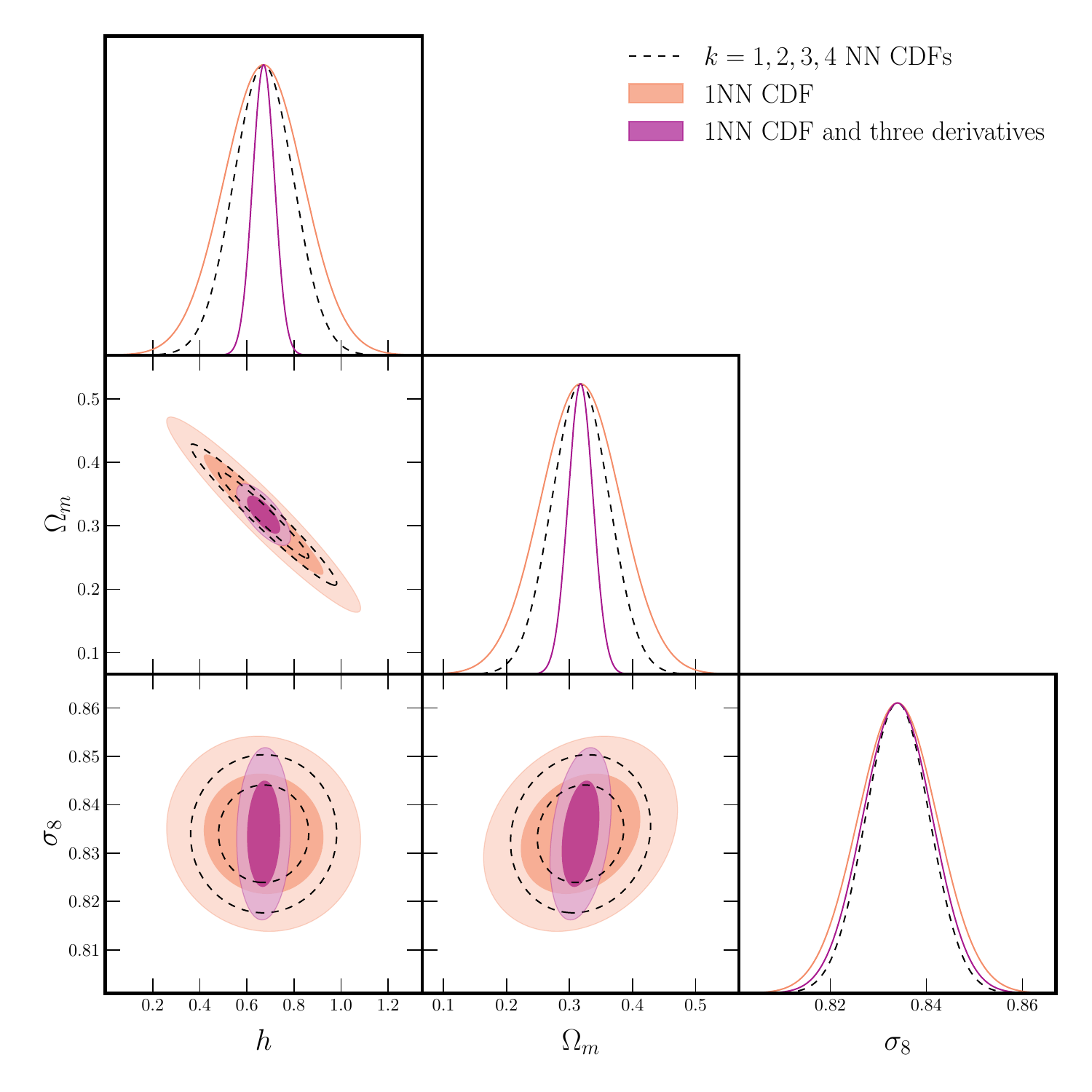}
    \caption{The data vectors are of the 1NN CDF at 16 logarithmic bins compared with the 1NN CDF and its first three derivatives, each at 16 logarithmic bins. For reference, the contours for $k$NN CDFs for $k$=1,2,3,4 are plotted with dashed lines. Note that adding the derivatives to the data vector increases its constraining power. Also note that the 1NN CDF and its derivatives perform better than $k=1,2,3,4$ CDFs for $\{ h, \Omega_m \}$.}
    \label{fig:1nncdf_vs_1nnder}
\end{figure}
For each $k$, the 16 elements are spaced logarithmically, starting from $r_{\mathrm{min}}$ to $r_{\mathrm{max}}$. To avoid running into the tails of the CDF where it is noisy due to the finite number of query points, we place additional cuts on $r$. Thus, for a given $k$:
\begin{equation*}
    \begin{aligned}
       & r_{\mathrm{min}} = \max(4\; h^{-1} \mathrm{Mpc},\; r_{0.025}); \\ 
       & r_{\mathrm{max}} = \min(40\; h^{-1} \mathrm{Mpc},\; r_{0.975})
    \end{aligned}
\end{equation*}
where $\mathrm{CDF}_{k\mathrm{NN}} (r_{0.025}) = 0.025$ and $\mathrm{CDF}_{k\mathrm{NN}} (r_{0.975}) = 0.975$. For the derivatives, we follow the $r$ bins for the corresponding CDF. For the Minkowski Functionals, we stick with the $r$ bins corresponding to $\mathrm{CDF}_{1\mathrm{NN}}$, since equation \ref{eq:W0_2} relates the two of them. We first compute the CDFs and $r$ bins for 100 fiducial realizations. We use the mean from 100 fiducial realizations to fix the radial bins where the CDFs are well-measured. These bins are then fixed through the analysis.
\\ \\
We numerically compute the derivatives of the CDFs by counting the query points within the volume of interest at radius $r, r\pm \epsilon, r\pm 2\epsilon, r\pm 3\epsilon$ and $r\pm 4\epsilon$ for a small $\epsilon$; and then combining these using the central finite difference method. This allows us to achieve a balance between speed and accuracy, although the method to compute the derivatives could be further optimized in future work. For computing the Minkowski Functionals, we use equations \ref{eq:mean_curv_arclength} and \ref{eq:chi_arclength} and perform appropriate calculations by isolating the query points that fall near the intersection regions of spheres. 
\begin{figure}
    \centering
    \includegraphics[width=\linewidth]{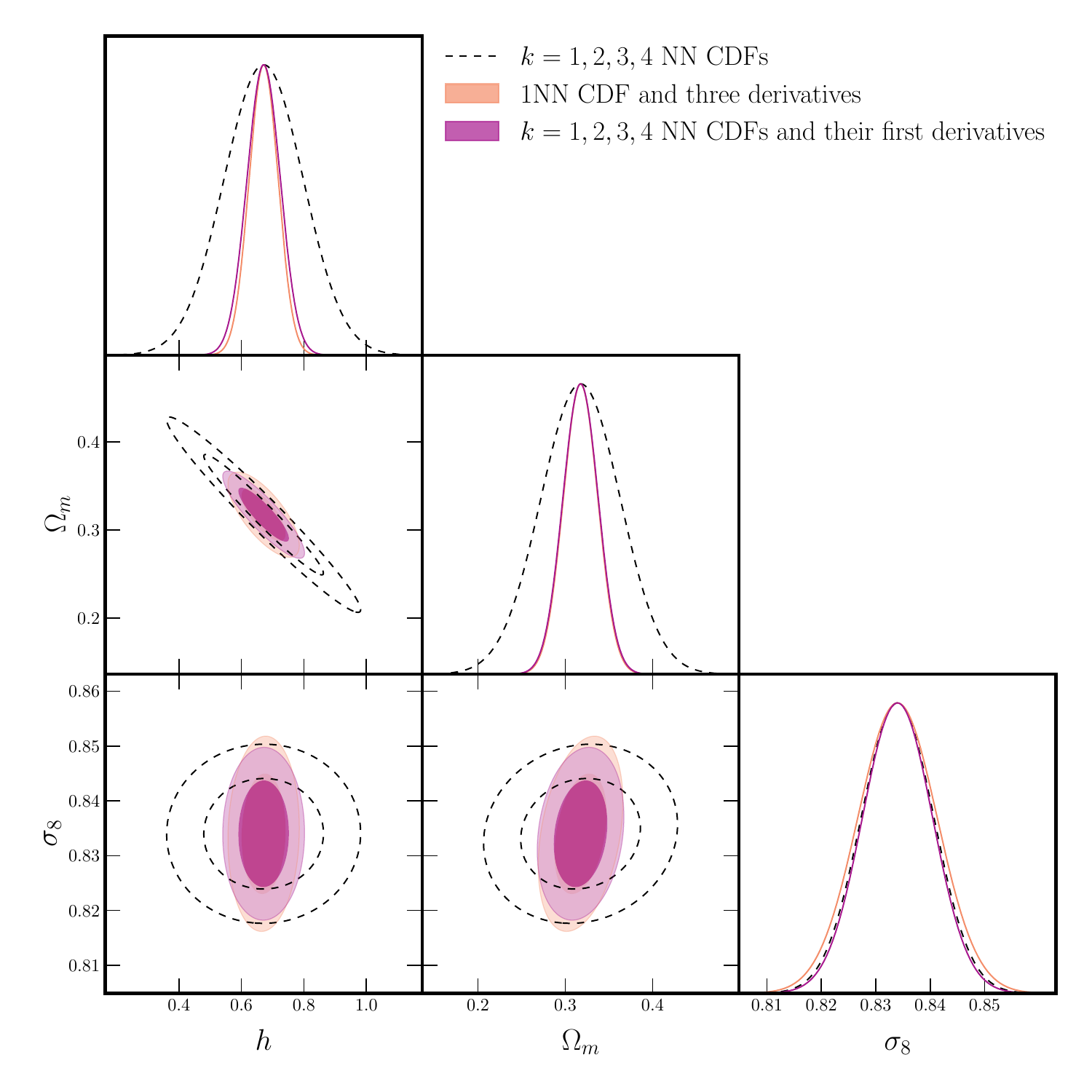}
    \caption{The data vectors are of the 1NN CDF and its first three derivatives, each at 16 logarithmic bins compared with the $k$NN CDFs and their first derivatives for $k=1,2,3,4$, each at 16 logarithmic bins. For reference, the contours for $k$NN CDFs for $k$=1,2,3,4 are plotted with dashed lines. In order to use the information in both higher $k$NN CDFs as well as derivatives of the CDFs, we created a data vector with $k=1,2,3,4$ NN CDFs and their first derivatives. This data vector gives us the best parameter constraints out of all the data vectors we evaluate in this paper.}
    \label{fig:1nnder_vs_knnder}
\end{figure}
\begin{figure}
    \centering
    \includegraphics[width=\linewidth]{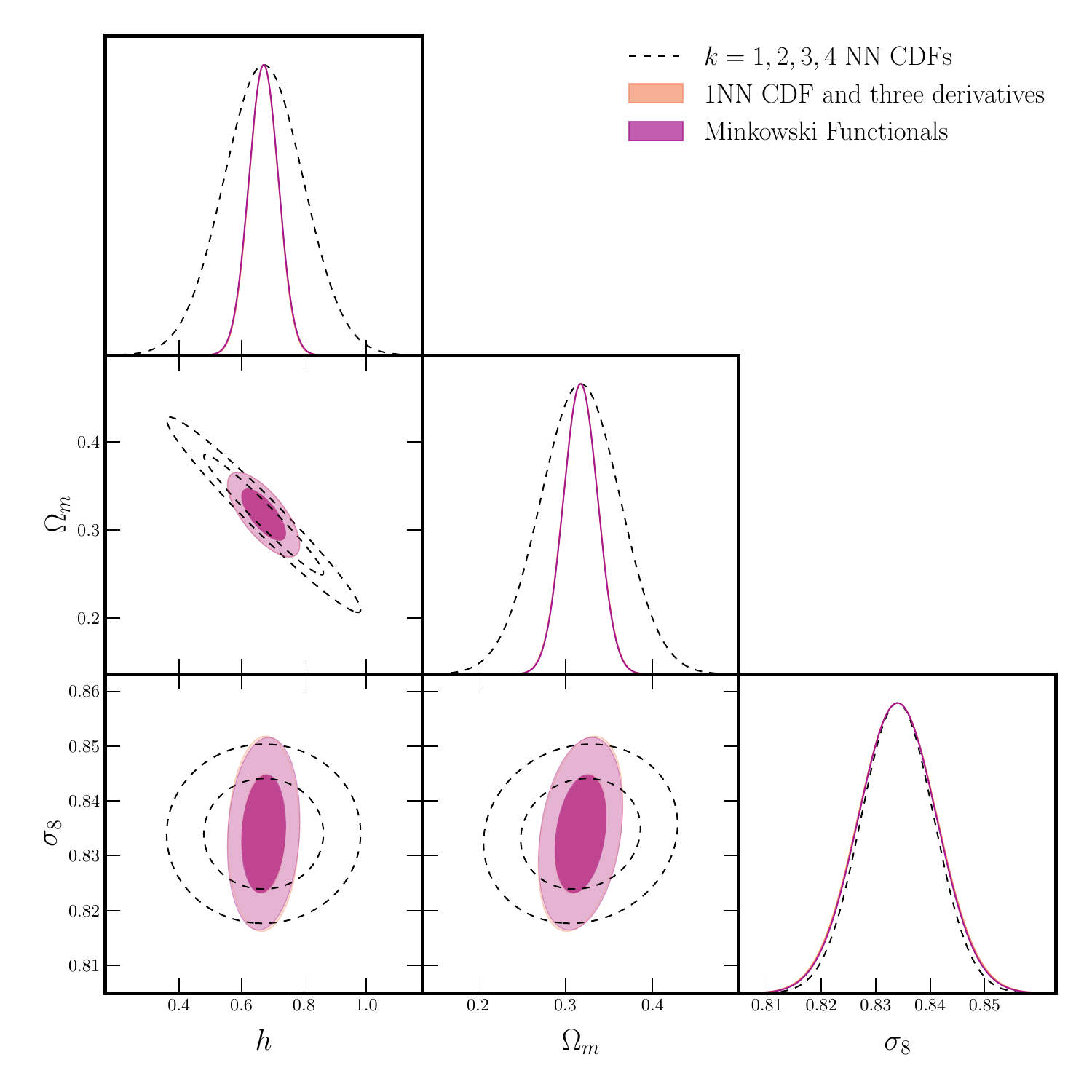}
    \caption{The data vectors are of the 1NN CDF and its first three derivatives, each at 16 logarithmic bins compared with the four Minkowski Functionals, each at 16 logarithmic bins. For reference, the contours for $k$NN CDFs for $k$=1,2,3,4 are plotted with dashed lines. We established a correspondence between the 1NN CDF and its derivatives and the Minkowski Functionals in seciton \ref{sec:knn_mink}. Here, we see nearly identical constraining power from both the statistics.}
    \label{fig:1nnder_vs_mink}
\end{figure}
\\ \\
The data vector for the 1NN CDF and its three derivatives, and their derivative with respect to the cosmological parameters are all plotted in figure \ref{fig:data_vectors}. Note that for the second, third, and fourth columns, the relative errors are high at some $r$ because there the data vector values are near $0$, so small fluctuations also lead to large relative errors. This is not a cause for concern, as the absolute errors remain fairly constant throughout, as shown in the topmost panel. In fact, upon removal of the points that have a near-zero value, the Fisher constraints increase. So we decide to keep these points in our analysis. 
\\ \\
Note the behaviour of $\mathrm{CDF}_{1\mathrm{NN}}$ and $\frac{d}{dr} \mathrm{CDF}_{1\mathrm{NN}}$ as a function of $r$.
The volume within the union of spheres of radius $r$ increases monotonically from zero at $r=0$ to the full box volume $V_0$ at sufficiently large $r$, which is reflected in the behaviour of $\mathrm{CDF}_{1\mathrm{NN}}(r)$. Similarly, the surface area of the union of spheres starts increasing from  zero at $r=0$, but soon the decrease in area due to merging of spheres dominates over the increase in area due to the expanding spheres, so the total surface area decreases, till all spheres have merged and the entire box is covered at a sufficiently large $r$. This can be seen in the behaviour of $\frac{d}{dr} \mathrm{CDF}_{1\mathrm{NN}}(r)$. Unlike the second and third derivative, $\mathrm{CDF}_{1\mathrm{NN}}$ and $\frac{d}{dr} \mathrm{CDF}_{1\mathrm{NN}}$ are always positive, as is expected from their geometric interpretation in terms of volumes and areas. 
\\ \\
The matrix $F_{\alpha\beta}$ is the covariance matrix assuming a multinormal probability distribution for the parameters $\alpha$ and $\beta$ around their fiducial values. Thus, we get a two-dimensional Gaussian distribution for every pair of parameters. We can compare the constraining power of various data vectors by comparing the variance of these Gaussians. We can even understand the constraining power of pairs of parameters from the two-dimensional Gaussians. This is particularly useful because different data vectors might have different constraining powers along different axes, and combining all of their information can give us the best joint constraints on the parameters. 
\begin{figure}
    \centering
    \includegraphics[width=\linewidth]{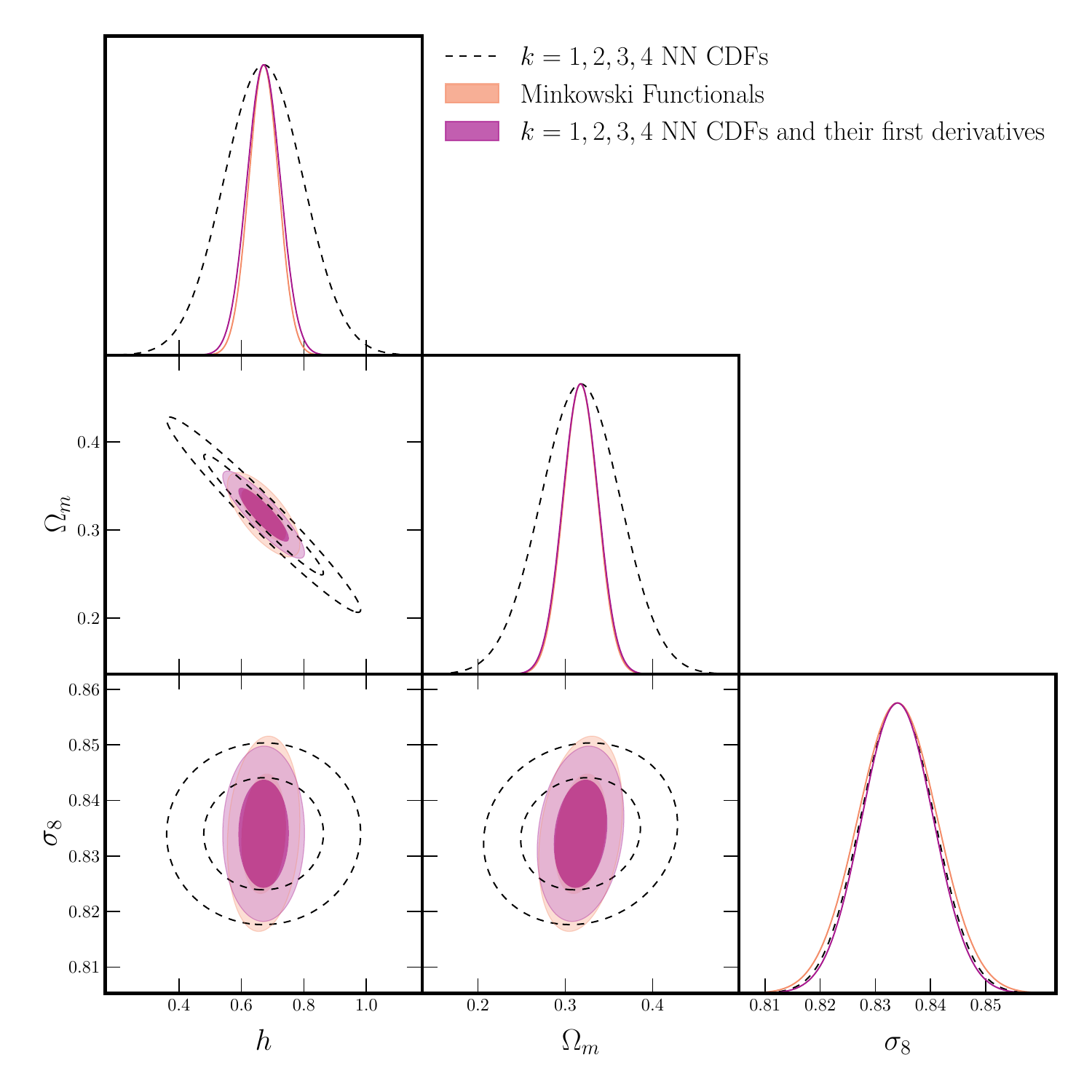}
    \caption{The data vectors are of the four Minkowski Functionals, each at 16 logarithmic bins compared with the $k$NN CDFs and their first derivatives for $k=1,2,3,4$, each at 16 logarithmic bins. For reference, the contours for $k$NN CDFs for $k$=1,2,3,4 are plotted with dashed lines. The $k=1,2,3,4$ NN CDFs and their first derivatives give tighter constraints than the Minkowski Functionals for all the cosmological parameters.}
    \label{fig:mink_vs_knnder}
\end{figure}
\begin{table*}
    \centering
    \caption{The $1\sigma$ constraints obtained on the cosmological parameters $\{h,\Omega_m, \sigma_8\}$ from our Fisher analysis.}
    \begin{tabular}{|l|c c c|}
    \hline
    Data Vector ($\mathbf{D}$) \hspace{9.5cm} $1\sigma$ constraints on & $h$ & $\Omega_m$ & $\sigma_8$\\
    \hline\hline 
    $\{\mathrm{CDF}_{1\mathrm{NN}}\}$ & 0.1684 & 0.0628 & 0.0082 \\ \hline 
    $\{\mathrm{CDF}_{1\mathrm{NN}}, \mathrm{CDF}_{2\mathrm{NN}}, \mathrm{CDF}_{3\mathrm{NN}}, \mathrm{CDF}_{4\mathrm{NN}}\}$ & 0.1268 & 0.0454 & 0.0067 \\ \hline 
    $\{\mathrm{CDF}_{1\mathrm{NN}}, \frac{d}{dr}\mathrm{CDF}_{1\mathrm{NN}}, \frac{d^2}{dr^2}\mathrm{CDF}_{1\mathrm{NN}}, \frac{d^3}{dr^3}\mathrm{CDF}_{1\mathrm{NN}}\}$ & 0.0467 & 0.0198 & 0.0073 \\ \hline 
    $\{\mathrm{CDF}_{1\mathrm{NN}}, \frac{d}{dr}\mathrm{CDF}_{1\mathrm{NN}}\}$ (equivalent to $\{W_0, W_1\}$) & 0.0918 & 0.0341 & 0.0077 \\ \hline 
    $\{\frac{d^2}{dr^2}\mathrm{CDF}_{1\mathrm{NN}}, \frac{d^3}{dr^3}\mathrm{CDF}_{1\mathrm{NN}}\}$ & 0.0799 & 0.0295 & 0.0141 \\ \hline 
    $\{\mathrm{CDF}_{1\mathrm{NN}}, \frac{d}{dr}\mathrm{CDF}_{1\mathrm{NN}}, \mathrm{CDF}_{2\mathrm{NN}}, \frac{d}{dr}\mathrm{CDF}_{2\mathrm{NN}}, \mathrm{CDF}_{3\mathrm{NN}}, \frac{d}{dr}\mathrm{CDF}_{3\mathrm{NN}}, \mathrm{CDF}_{4\mathrm{NN}}, \frac{d}{dr}\mathrm{CDF}_{4\mathrm{NN}}\}$ & 0.0535 & 0.0203 & 0.0064 \\ \hline 
    $\{W_0, W_1, W_2, W_3\}$ & 0.0474 & 0.0196 & 0.0072 \\ \hline 
    $\{W_2, W_3\}$ & 0.0614 & 0.0231 & 0.0095 \\ \hline 
    \hline
\end{tabular}
    \label{tab:fisher}
\end{table*}
\\ \\
From figure \ref{fig:1nncdf_vs_1nnder} we infer that adding the information of the derivatives significantly improves the constraints, even though theoretically the information about the derivatives should be present in the full function itself. This is because we are sampling the CDF at 16 bins only, and this removes the information of how the function changes at those points, which the derivative captures. Thus, compressing the CDF into a data vector with 16 bins is not an optimal choice. It should be noted that computing the derivatives of the CDF does not add significantly to the computing time, as performing the $k$-d tree algorithm on query points is the most time consuming step and only needs to be done once. We also see that the constraining power of the 1NN CDF and its derivatives is comparable to the $k$NN CDFs for $k=1,2,3,4$ for $\sigma_8$, and the 1NN CDF and derivatives perform better for $\{h,\Omega_m\}$. This is an illuminating result, as it suggests that adding derivatives of the 1NN CDF provides more information than incorporating higher-order CDFs. This implies that the data vectors of the CDF alone fail to capture a significant amount of information that the full shape of the CDF theoretically contains. Since the derivatives do not take much additional time to compute, incorporating them into the data vectors is a quick way to achieve better constraining power.
\\ \\
We compared the relative contribution of the 1NN CDF and its first derivative versus the second and third derivative, and the role of the former was slightly more in constraining the cosmological parameters, especially for $\sigma_8$. In table \ref{tab:fisher} we can see roughly a factor of 2 improvement in the constraining power of $\sigma_8$ for the CDF and first derivative, compared to the second and third derivative, while the other parameters fare similarly. In order to take advantage of using both higher $k$'s and information in the derivatives, we use the $k$NN CDFs for $k=1,2,3,4$ and their first derivatives as our new data vector. Note that even though the second and third derivatives constrain $\{h,\Omega_m\}$ slightly better, long data vectors slow down the analysis, as well as increase demands on the number of simulations needed to accurately calibrate the covariance matrix through the Hartlap factor (equation \ref{eq:hartlap}). Thus, in our new data vector we only keep the CDFs and their first derivatives. From figure \ref{fig:1nnder_vs_knnder}, we conclude that adding just the first derivatives leads to a significant improvement in constraints compared to only the $k=1,2,3,4$ NN CDFs. However, these constraints are very similar to those obtained from Minkowski Functionals and 1NN CDF and derivatives, with only marginal differences (see figures \ref{fig:1nnder_vs_knnder} and \ref{fig:mink_vs_knnder}). 
\\ \\ 
Since each successive $k$-nearest neighbour CDF looks at particles farther away on average, 1NN CDF best captures the information on the smallest scales and higher $k$NN CDFs are more informative on larger scales. Thus, having a larger value of $r_{\mathrm{min}}$ will transfer some of the constraining power of lower $k$NN CDFs into the higher $k$NN CDFs, altering the constraints slightly. All the results quoted in this paper are for $r_{\mathrm{min}}=4\; h^{-1} \mathrm{Mpc} $. But upon increasing $r_{\mathrm{min}}$, the $k$NN CDFs and their derivatives become slightly more constraining compared to the 1NN CDFs and their derivatives, or the Minkowski Functionals. The choice ultimately depends on the scales of interest for the analysis.
\\ \\
We showed the correspondence between the 1NN CDF and its derivatives and Minkowski Functionals in section \ref{sec:knn_mink}, thus expecting similar constraining power using both data vectors. That is what we observe in figure \ref{fig:1nnder_vs_mink}. From table \ref{tab:fisher} we further infer that the data vector containing $W_2, W_3$ has a better constraining power than the second and third derivative of the 1NN CDF. Even though both have similar information and sensitive to the same geometric features in the dataset, we hypothesize that the Minkowski Functionals provide that information in a slightly more accessible way. For example, to extract the Euler characteristic (or $W_3$) from the third derivative of 1NN CDF, we need the CDF with infinite precision. However, the combined data vector of the 1NN CDF and its three derivatives gives constraints similar to those from the four Minkowski functionals combined. The CDF and derivatives provide these constraints with just 16 discrete bins, while requiring lesser computation time. This makes them a good choice for scaling up to larger datasets.
\\ \\ 
Computing Minkowski Functionals has been a very slow process, and thus been used for relatively small datasets in the literature so far. However, using $k$NN CDFs and their derivatives gives us constraints competitive to the Minkowski Functionals but much faster. In fact, for our analysis we computed the Minkowski Functionals using the $k$-d tree $k$NN algorithm applied on query points. This algorithm speeds up the calculation of the Minkowski Functionals significantly compared to previous methods. The algorithm for the same will be explained in an upcoming paper. 
\\ \\
Table \ref{tab:fisher} summarizes the $1\sigma$ constraints we get on the cosmological parameters from all the data vectors we have used in our Fisher analysis. 
\\ \\
To sum up, from the Fisher analysis we found that:
\begin{enumerate}
    \item Adding the derivatives to the $k$NN CDF data vectors significantly increases the constraining power, suggesting that just the CDFs sampled at a few values of $r$ is missing out on a lot of information. This information can be  accessed by adding the derivatives into the data vector. \\
    \item We showed in section \ref{sec:knn_mink} that the 1NN CDF and its three derivatives contain similar information to the Minkowski Functionals. Using a Fisher analysis, we conclude the same. The CDFs can be computed much faster, making them more useful for scaling up to large datasets from cosmological surveys. \\
    \item Combining the strategy of using higher $k$’s and derivatives, we construct a data vector comprising the $k = 1, 2, 3, 4$ NN CDFs and their first derivatives. On the scales considered in our analysis, this data vector performs comparably to the Minkowski Functionals and the 1NN CDF with derivatives, without offering a significant improvement. However, for larger scales, the $k$NN CDFs and their derivatives tend to become more constraining than the Minkowski Functionals and the 1NN CDF with derivatives.
\end{enumerate}
Looking forward, we need to either use a combination of the $k$NN CDFs and their derivatives, or find better ways to compress the information in the CDFs without losing much constraining power, in order to obtain stronger bounds using the $k$NN CDFs.

\section{Summary and Discussion}
\label{sec:final}
The $k$-nearest neighbour cumulative distribution functions are measures of clustering for cosmological datasets that are sensitive to non-Gaussianities at small scales. They are fast to compute, and can be used to study a wide range of cosmological datasets, from discrete data to continuous data and even cross-correlations between these. In this paper, we developed geometric interpretations for the $k$NN CDFs, their derivatives, and the joint cross-correlation CDFs. We demonstrated that the 1NN CDF represents the fraction of volume contained within spheres of equal radius centered on data points. We also presented a geometric interpretation for higher $k$NN CDFs ($k>1$) in terms of volume within intersections of multiples spheres, and cross-correlation CDFs in terms of volume within intersections of $k$NN regions for the two sets of data points. 
\\ \\
We mathematically demonstrated how the derivatives of the 1NN CDF are sensitive to geometric properties such as surface area of the spheres; and the angles, solid angles, and arcs formed at the intersections of spheres around data points. The Minkowski Functionals are also sensitive to the same quantities, albeit in a different combination. We established a correspondence between the 1NN CDF, its three derivatives, and the four Minkowski Functionals, highlighting that both encapsulate equivalent information and thus possess comparable constraining power.
\\ \\
We then did a series of Fisher comparisons between many different data vectors that use the $k$NN CDFs, their derivatives, and the Minkowski Functionals. Our findings indicate that the current approach of using 16 bins to summarize the CDFs is suboptimal. Incorporating the information from the derivatives of these CDFs significantly enhances their constraining power. In fact, using just the 1NN CDF and its three derivatives gave us better constraints than using $k=1,2,3,4$ NN CDFs. This signals the need to devise improved methods for summarizing the $k$NN CDFs to achieve tighter parameter constraints.
\\ \\
One possible way to obtain tighter constraints is to use more bins and incorporate the derivatives of the CDFs. However, arbitrarily increasing the number of bins leads to a more imperfect measurement of the (co)variance, which can be seen from the dependence of the Hartlap factor (equation \ref{eq:hartlap}) on the number of elements in the data vector, $p$. On the other hand, having too few bins leads to poorer constraints as it misses out on information present in the full shape of the CDF. One possible avenue for future research is to determine the optimal length of the $k$NN CDF data vectors. Additionally, other methods to summarize the full shape of the CDF can also be investigated. For example, \cite{refId1} provide a parametric model to describe the $k$NN CDFs. Although parameterizing the CDFs is a promising approach, it is crucial to ensure that the parametric model accurately represents the CDF and its derivatives across a wide range of length scales for practical applicability.
\\ \\
After establishing a correspondence between the 1NN CDF, its derivatives, and the Minkowski Functionals, we compared their constraining power using Fisher analysis. As expected, we find that both have very similar Fisher constraints. In order to combine the strengths of using higher $k$'s as well as derivatives of the CDFs, we created a data vector containing the $k=1,2,3,4$ NN CDFs and their first derivatives. This data vector performed comparably to the 1NN CDF and derivatives, and the Minkowski Functionals.
\\ \\
The computational efficiency of the CDFs makes them highly scalable, facilitating their application to large datasets from stage 4 galaxy surveys. Additionally, unlike the Minkowski Functionals the $k$NN CDFs can be naturally extended to study cross-correlations. This is especially relevant because cross-correlations are becoming increasingly important in cosmology. For example, cross-correlations can be used to mitigate noise present in autocorrelations as well as to detect cosmological fields otherwise drowned in large foreground (e.g., \cite{10.1111/j.1745-3933.2008.00581.x}, \cite{2016PhRvL.117e1301H}, \cite{Amiri_2024}).
\\ \\
Moreover, we computed the Minkowski Functional data vectors using various computations on the query point positions after performing the $k$-d tree algorithm. More on this and other algorithms that make use of the spatial information of $k$NN query points will be discussed in a future paper. The main advantage of using the $k$-d tree algorithm is that it only needs to be performed once to obtain the relevant statistic at all scales. Going from applying the $k$-d tree algorithm and sorting the query points to constructing a CDF from these distances results in information compression. This formulation makes the $k$NN CDFs more tractable, as they are solely functions of the length scale  $r$. However, this compression may result in the loss of information that is more effectively captured by topological methods such as Persistent Homology or Betti Numbers. This suggests potential avenues for extracting information directly from the $k$NN distance maps, prior to their compression into CDFs, to achieve constraints comparable to topological data analysis methods. This is currently being investigated using machine learning methods (\cite{ChatterjeeInPrep}). Furthermore, comparisons between the constraining power of $k$NN CDF and Betti Curves are also being investigated (\cite{BettikNN}). 
\\ \\
The connections between the many higher-order statistics used to characterize non-Gaussian clustering ($N$-point correlation functions, VPFs and CICs, $k$NN CDFs) was discussed in \cite{10.1093/mnras/staa3604}. Our work explores the geometric information contained in the $k$NN CDFs, and its connections with the Minkowski Functionals. The correspondence between the $k$NN CDFs and Minkowski Functionals, despite originating from fundamentally different approaches to characterizing clustering, is particularly noteworthy. With this work, we now have a common framework relating many of the higher-order clustering statistics, including geometric measures of clustering. Investigating the relationships among different statistics is crucial for identifying the specific aspects of clustering that each statistic effectively captures. These insights will be crucial in designing analysis pipelines for stage 4 cosmological surveys. 

\section*{Acknowledgements}
The authors thank the suggestions given by the anonymous referee to help improve the paper. AB thanks Pratyush Pranav for helpful discussions in the initial stages of the work. The authors thank Gilbert Holder, Pravabati Chingangbam, Kaustubh Rajesh Gupta and Susmita Adhikari for helpful suggestions on the manuscript. AB’s work was partially supported by the Startup Research Grant (SRG/2023/000378) from the Science and Engineering Research Board (SERB), India. This work was also supported by U.S. Department of Energy grant DE-AC02-76SF00515 to SLAC National Accelerator Laboratory managed by Stanford University. The authors acknowledge the PARAM Brahma Facility under the National Supercomputing Mission, Government of India, at the Indian Institute of Science Education and Research, Pune, for providing the computing resources for this work. The Python libraries \textsc{NumPy}\footnote{\url{https://numpy.org/}} and \textsc{SciPy}\footnote{\url{https://scipy.org/}} were used extensively in this work. All plots in this paper were generated using \textsc{Matplotlib}\footnote{\url{https://matplotlib.org/}}, and all geometric figures were made using \textsc{TikZ}\footnote{\url{https://github.com/pgf-tikz/pgf}}. The authors acknowledge the use of \textsc{ChatGPT} for refining the text.

\section*{Data Availability}
The simulation data used in this paper is publicly avail-
able at \url{https://quijote-simulations.readthedocs.io/en/latest/access.html}. The data generated during this work is available on reasonable request.
\bibliographystyle{mnras}
\bibliography{references.bib} 
\appendix
\section{First and second derivative of volume within spheres}
\label{sec:arc}
\begin{figure}
    \centering
    \includegraphics[width=\linewidth]{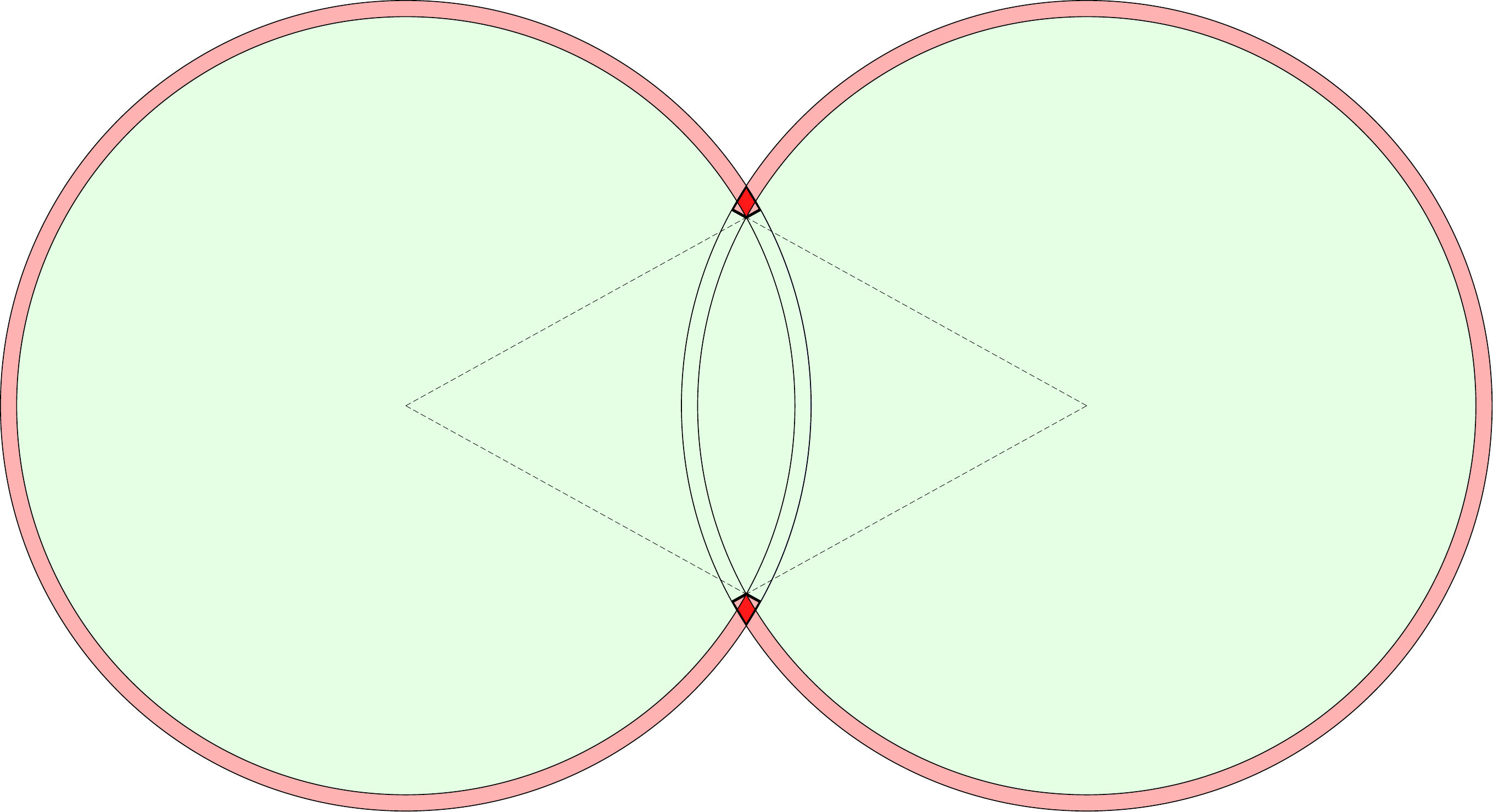}
    \caption{$V(r+\Delta r)$ expressed as a sum; as written in equation~\ref{eq:rplusdelr}. $V(r)$ is represented by the green region, $\frac{A(r)}{4\pi r^2}\frac{4\pi}{3}\left( (r+\Delta r)^3 - r^3\right) $ is represented by the red region (the dark red part is added twice), and $V_{\mathrm{arc}}$ is represented by the region within the thick black outline. The figure shows the simple case of two spheres, but equation~\ref{eq:rplusdelr} is true even for more spheres (neglecting contribution due to points of intersection of three or more spheres).}
    \label{fig:rplusdelr}
\end{figure}

\begin{figure}
    \centering
    \includegraphics[width=0.9\linewidth]{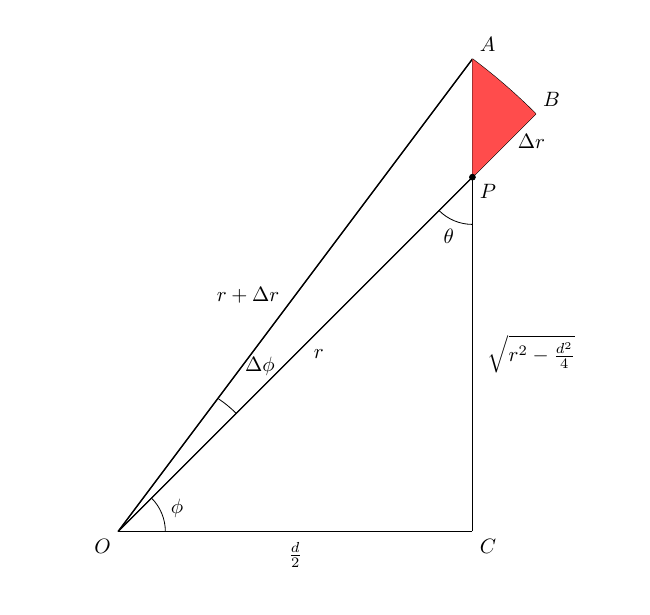}
    \caption{Half of the cross sectional area of the toroid that contributes to $V_{\mathrm{arc}}$. The full volume is made by reflecting the red figure about the vertical line AP, and then rotating it $360 ^\circ$ around the center C and axis OC.}
    \label{fig:cross_section}
\end{figure}

Let the volume within the union of spheres of radius $r$ be $V(r)$, and the surface area be $A(r)$. Then, (see figure~\ref{fig:rplusdelr})
\begin{equation}
    V(r+\Delta r) \approx V(r) + \frac{A(r)}{4\pi r^2}\frac{4\pi}{3}\left( (r+\Delta r)^3 - r^3\right) - V_{\mathrm{arc}} 
    \label{eq:rplusdelr}
\end{equation}
where we have neglected the smaller terms coming due to points where three or more spheres intersect. The second term with $A(r)$ quantifies the volume of the shells of thickness $\Delta r$ that arises when we simply extend the exposed area $A(r)$ outwards by a thickness $\Delta r$. This leads to intersections, which we remove by considering $V_{\mathrm{arc}}$.
\\ \\
To calculate $V_{\mathrm{arc}}$, we need to consider it as a sum of all the partial toroidal volumes formed by the intersection of pairs of spheres. The partial toroids might be formed due to a circular arc (with length $\ell_{ij}$) that is not a full circle (when more than 2 spheres interset). Let us consider just the contribution from one complete circle first, then we can add all the contributions with an additional weight factor to take into account the partial arcs. Since we need the second derivative, we will only keep terms till $(\Delta r)^2$ in all our series expansions. 
\newline\newline Using the geometry of the figure (\ref{fig:cross_section}), we know that 
\begin{equation}
    \cos\phi = \frac{d}{2r}\hspace{1mm},\hspace{10mm}
    \hspace{1cm} \sin\phi = \sqrt{1-\left(\frac{d}{2r}\right)^2}
\end{equation}
Taking the differentials for the first equation, we get
\begin{equation}
     -\sin\phi \Delta \phi = - \frac{d}{2r^2} \Delta r \hspace{1mm} \Rightarrow \hspace{1mm} \Delta \phi = \frac{d}{r\sqrt{4r^2-d^2}}\Delta r 
\end{equation}
Area of the sector $\mathrm{OAB}$ is
\begin{equation}
\begin{aligned}
    \mathrm{area(OAB)} &= \frac{\Delta \phi}{2}(r+\Delta r)^2 
    \\ &= \frac{d}{2r\sqrt{4r^2-d^2}}\left( r^2\Delta r + 2r(\Delta r)^2 + (\Delta r)^3 \right)
\end{aligned}
\end{equation}
Area of the triangle $\mathrm{OAP}$ is
\begin{equation}
\begin{aligned}
    \mathrm{area(OAP)} &= \frac{1}{2}r(r+\Delta r)\sin{\Delta \phi}
    \\ &= \frac{\Delta \phi}{2} (r^2 + r\Delta r) \left[ 
1 - \frac{(\Delta \phi)^2}{6} + \cdots \right]
\end{aligned}
\end{equation}
where we have used the Taylor expansion for $\sin{\Delta \phi}$. Thus, area of the region $\mathrm{PAB}$ (upto second order in $\Delta r$) is
\begin{equation}
\begin{aligned}
    \mathrm{area(PAB)} &= \mathrm{area(OAB)} - \mathrm{area(OAP)}
    \\& = \frac{d}{\sqrt{4r^2 - d^2}}(\Delta r)^2 
\end{aligned}
\end{equation}
To compute the volume of the toroid, we need to multiply twice of this area by the circumference of the circle with radius $r'$, where $r'$ is the distance from the midpoint of the two data points to the center of mass of the shape $\mathrm{APB}$ (Pappus's centroid theorem). Assume the center of mass radius to be
\begin{equation}
    r' = \sqrt{r^2 - \frac{d^2}{4}} + g(r,\phi)\Delta r
\end{equation}
Thus, keeping only upto second order terms in $\Delta r$ we get
\begin{equation}
\begin{aligned}
    (V_{\mathrm{arc}})_{ij} &= \left(2\pi r'\right)\times f_{ij}\times 2\left(\mathrm{area(PAB)} \right) \\
    &= f_{ij} \pi d (\Delta r)^2
\end{aligned}
\end{equation}
where $f_{ij}$ is the fraction of the complete circle that appears on the surface. Thus, 
\begin{equation}
    \ell_{ij} = 2\pi f_{ij} r \cos{\theta_{ij}}
\end{equation}
So the total summation of all these contributions becomes
\begin{equation}
\begin{aligned}
    V_{\mathrm{arc}} &= \frac{1}{2}\sum_{i,j} (V_{\mathrm{arc}})_{ij} 
    \\ &= \frac{\pi}{2} (\Delta r)^2 \sum_{i,j} f_{ij} d_{ij}
    \\ &= \frac{1}{2} (\Delta r)^2 \sum_{i,j} 2\pi f_{ij} r \sin{\theta_{ij}}
    \\ &= \frac{1}{2}(\Delta r)^2 \sum_{i,j} \ell_{ij} \tan{\theta_{ij}}
\end{aligned}
\label{eq:v_arc}
\end{equation}
where we have included a factor of $\frac{1}{2}$ before the summation to correct for the double counting of pairs of points. Thus, we can rewrite equation~\ref{eq:rplusdelr} using equation~\ref{eq:v_arc} (upto second order in $\Delta r$) as
\begin{equation}
    V(r+\Delta r) = V(r) + A(r)\Delta r + \left[\frac{A(r)}{r} - \frac{1}{2} \sum_{i,j} \ell_{ij} \tan{\theta_{ij}}\right] (\Delta r)^2
    \label{eq:rplusdelr2}
\end{equation}
Comparing it with the Taylor series upto second order in $\Delta r$: \newline $V(r+\Delta r) \approx V(r) + \frac{dV}{dr}\Delta r + \frac{d^2V}{dr^2}\frac{(\Delta r)^2}{2}$ and comparing this with equation~\ref{eq:rplusdelr2}, we get that
\begin{equation}
    A(r) = \frac{dV(r)}{dr}
\end{equation}
and, more importantly, 
\begin{equation}
    \frac{d^2V(r)}{dr^2} = 2 \left[\frac{A(r)}{r} - \frac{1}{2}\sum_{i,j}\ell_{ij}\tan{\theta_{ij}}\right]
    \label{eq:der2_arclength_appendix}
\end{equation}
\section{Minkowski Functionals: definition and calculation}
\label{sec:v_eps}
\begin{figure}
    \centering
    
    \begin{subfigure}[t]{0.45\columnwidth}
        \centering
        \includegraphics[width=0.7\columnwidth]{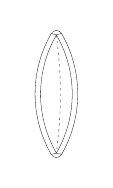}
        \caption{The $\epsilon$-volume $V_\epsilon(R)$ of the lens $S_1 \cap S_2$, where $S_1, S_2$ are spheres of radius $R$}
        \label{fig:alens}
    \end{subfigure}
    \hfill
    \begin{subfigure}[t]{0.5\columnwidth}
        \centering
        \includegraphics[width=0.63\columnwidth]{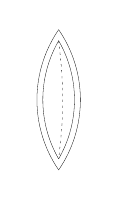}
        \caption{The lens formed by intersection of spheres of radius $R+\epsilon$.}
        \label{fig:blens}
    \end{subfigure}
    
    \caption{To calculate the Minkowski Functionals, we use~\ref{fig:alens}. But~\ref{fig:blens} is more appropriate when using $k$NN CDFs.}
    \label{fig:epsilon_vol_lens}
\end{figure}

\begin{figure}
    \centering
    
    \begin{subfigure}[t]{0.9\columnwidth}
        \centering
        \includegraphics[width=\columnwidth]{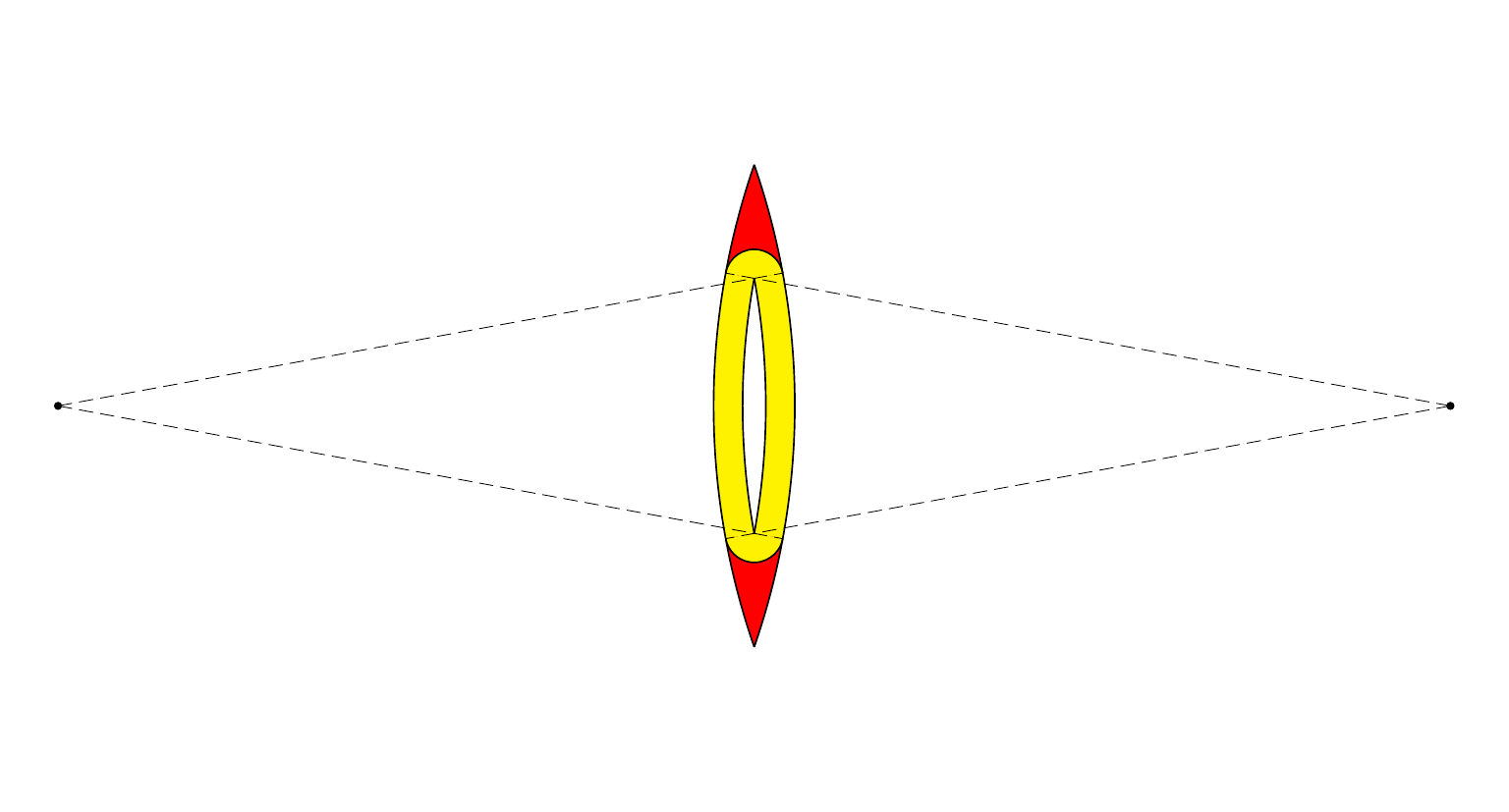}
        \caption{The lens when the radius of the spheres is small (the intersection has just happened).}
        \label{fig:a}
    \end{subfigure}
    \hspace{0.5\textwidth}
    \begin{subfigure}[t]{0.7\columnwidth}
        \centering
        \includegraphics[width=\columnwidth]{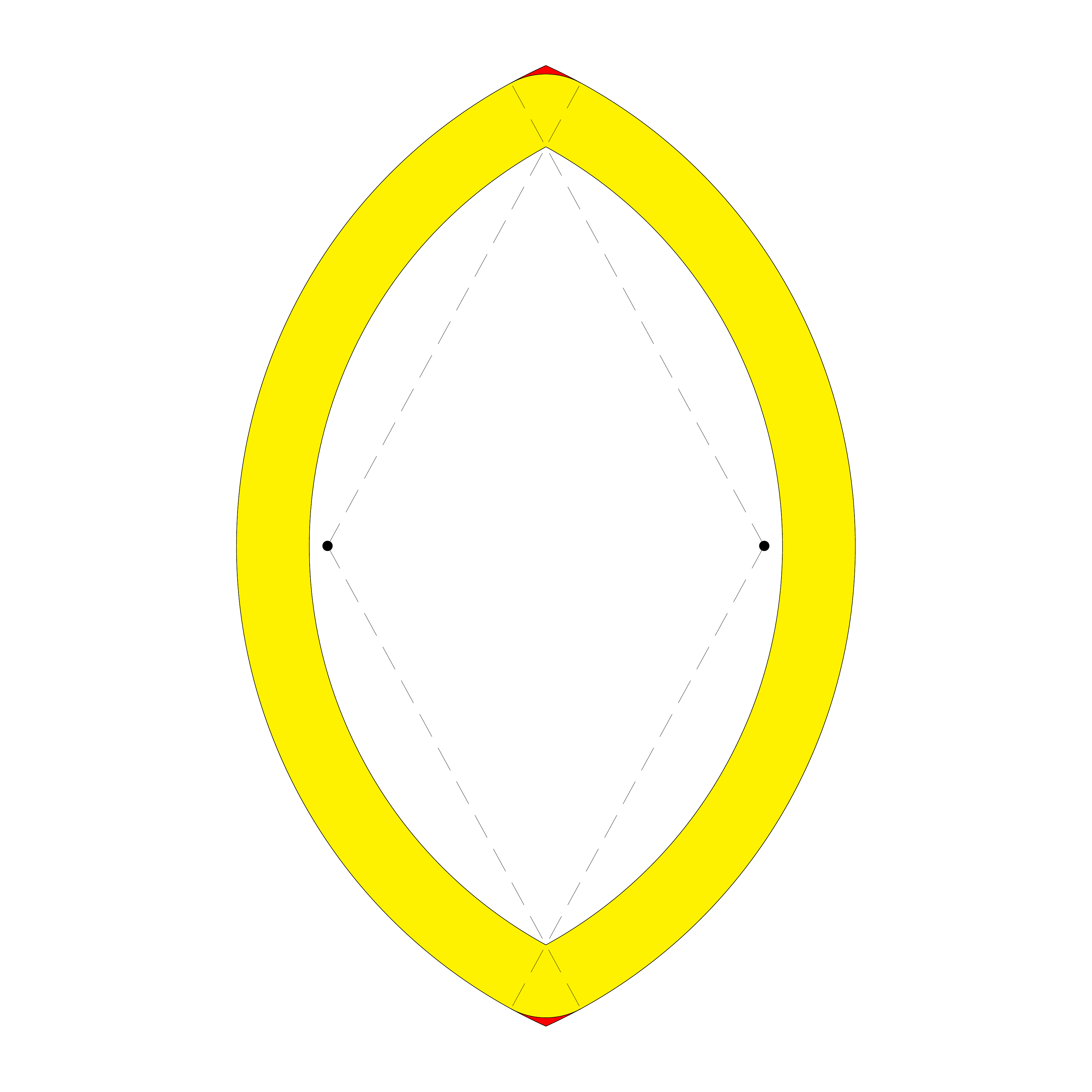}
        \caption{The lens when the radius of the spheres is much larger than the distance between the points.}
        \label{fig:b}
    \end{subfigure}
    
    \caption{The yellow region represents the correct $\epsilon$ region around the lens, with volume $V_\epsilon - V_{\mathrm{lens}}$. The red region represents the extra volume that is present when considering $V(r+\epsilon)$.}
    \label{fig:error}
\end{figure}

\begin{figure*}
    \centering
    \begin{subfigure}[t]{\textwidth}
        \centering
        \includegraphics[width=\textwidth]{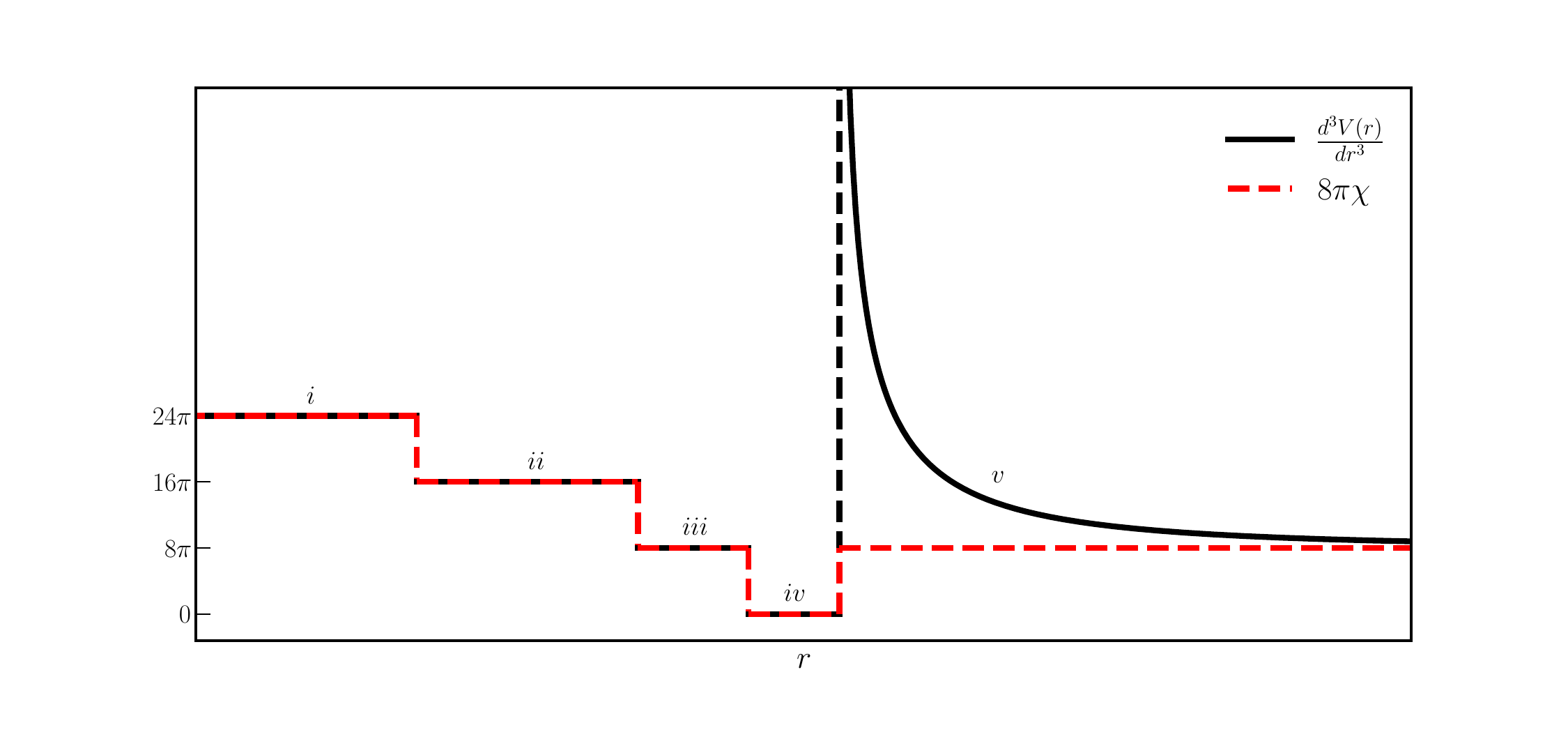}
        \caption*{}
    \end{subfigure}
    
    \vspace{-0.5cm} 
    
    \begin{subfigure}[t]{0.3\textwidth} 
        \centering
        \includegraphics[width=\textwidth]{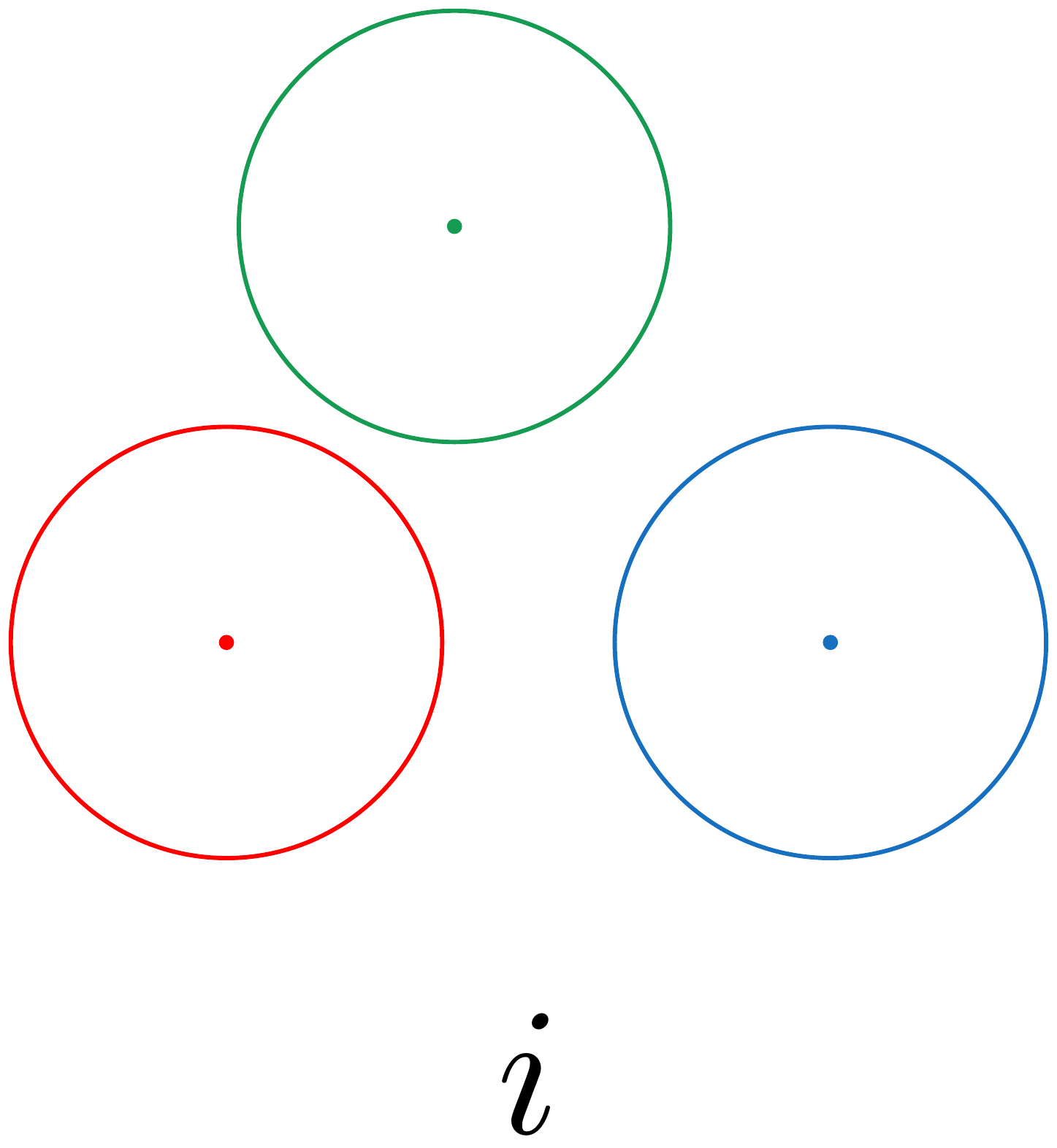}
        \caption*{}
    \end{subfigure}
    \hfill
    \begin{subfigure}[t]{0.3\textwidth}
        \centering
        \includegraphics[width=\textwidth]{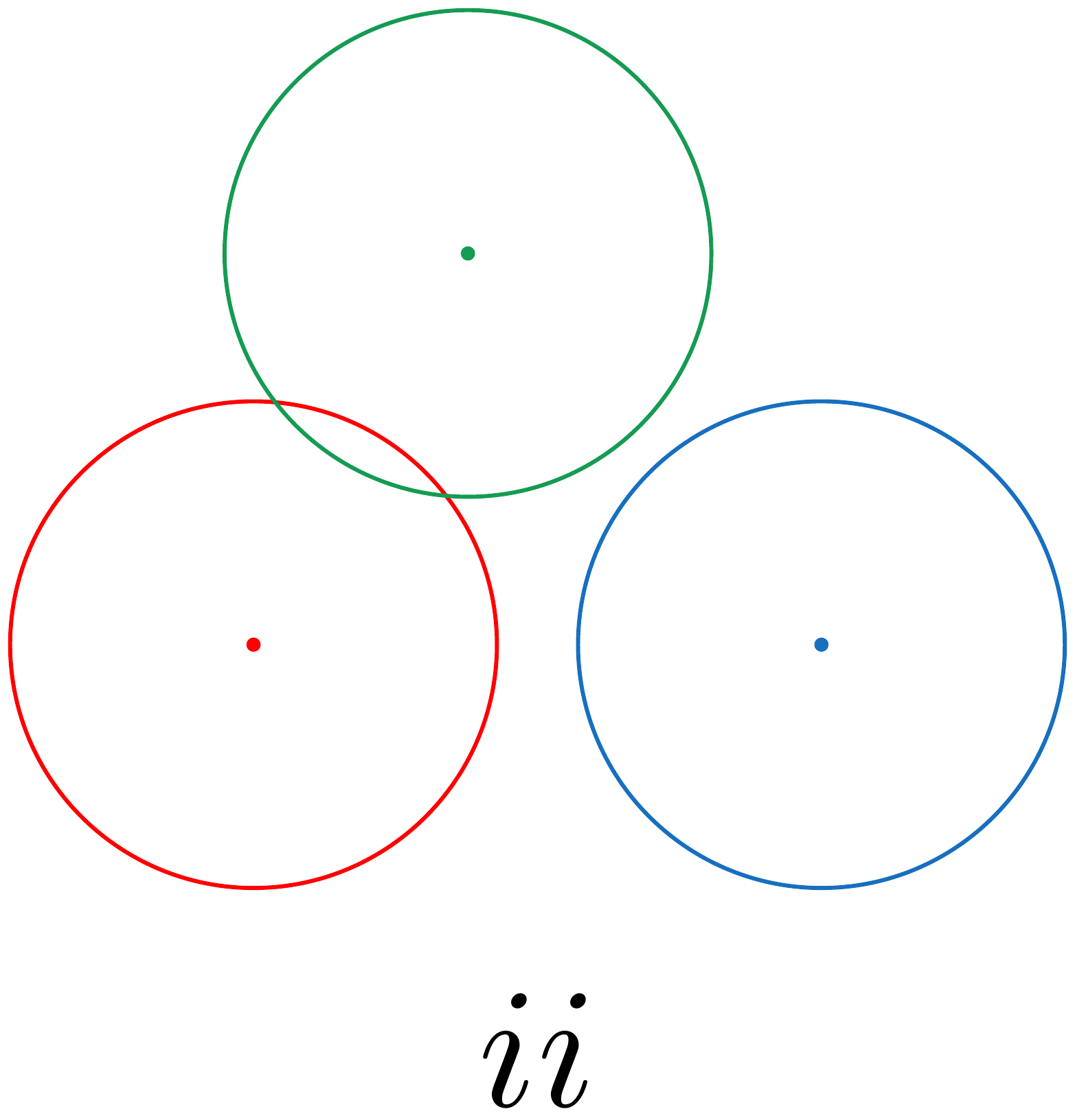}
        \caption*{}
    \end{subfigure}
    \hfill
    \begin{subfigure}[t]{0.3\textwidth}
        \centering
        \includegraphics[width=\textwidth]{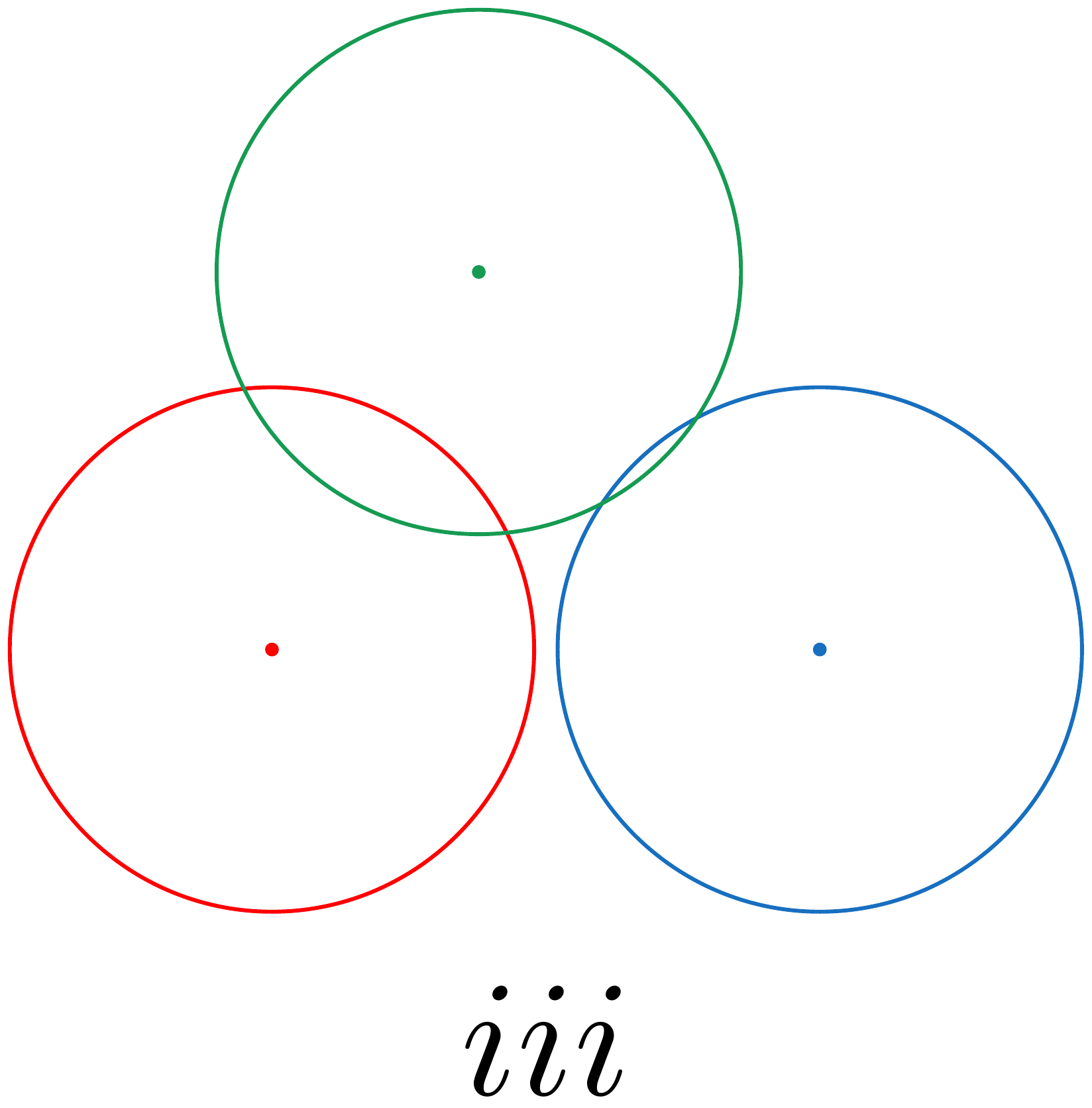}
        \caption*{}
    \end{subfigure}
    \hfill
    \begin{subfigure}[t]{0.3\textwidth}
        \centering
        \includegraphics[width=\textwidth]{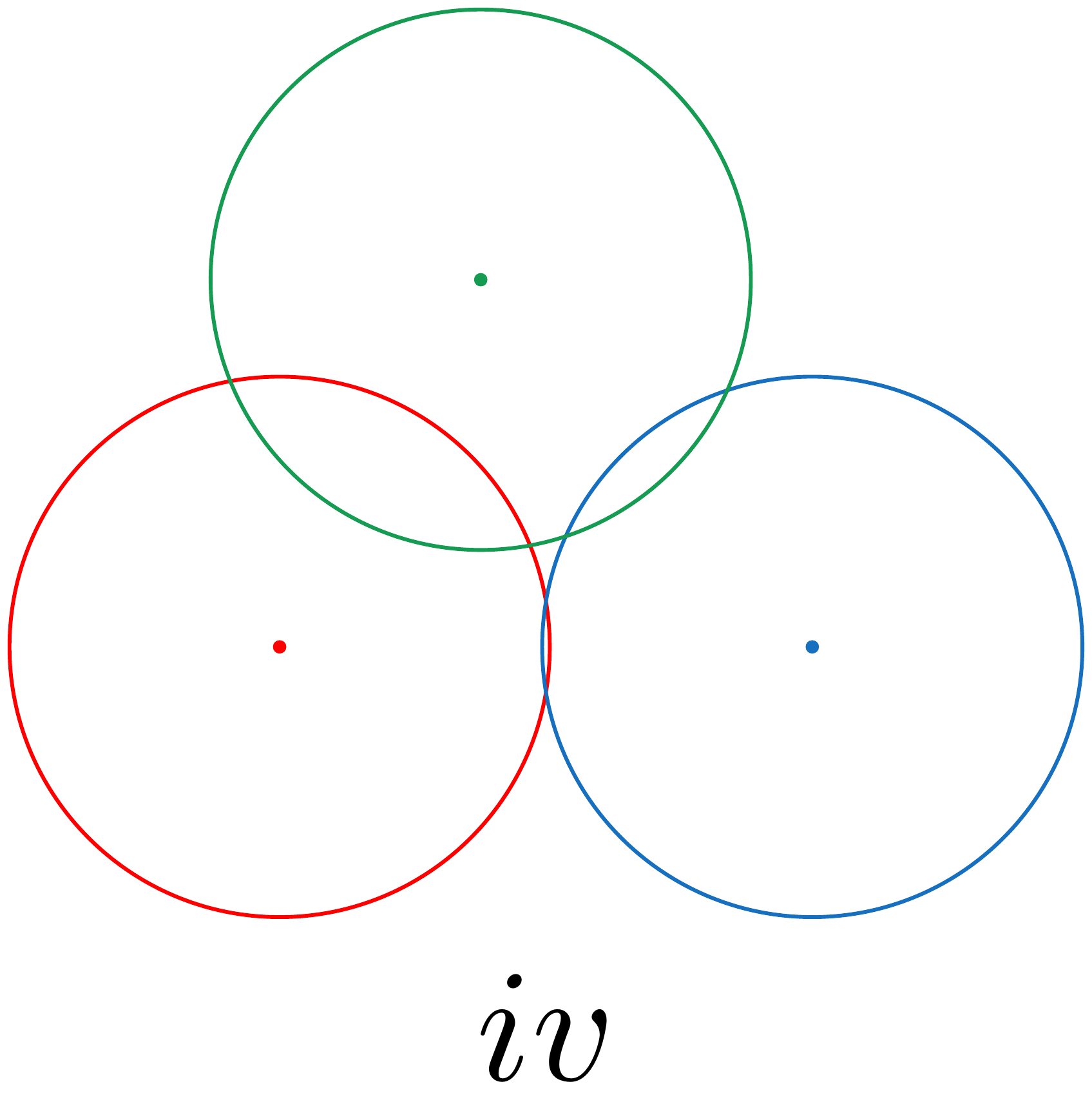}
        \caption*{}
    \end{subfigure}
    \hspace{0.1\textwidth} 
    \begin{subfigure}[t]{0.3\textwidth}
        \centering
        \includegraphics[width=\textwidth]{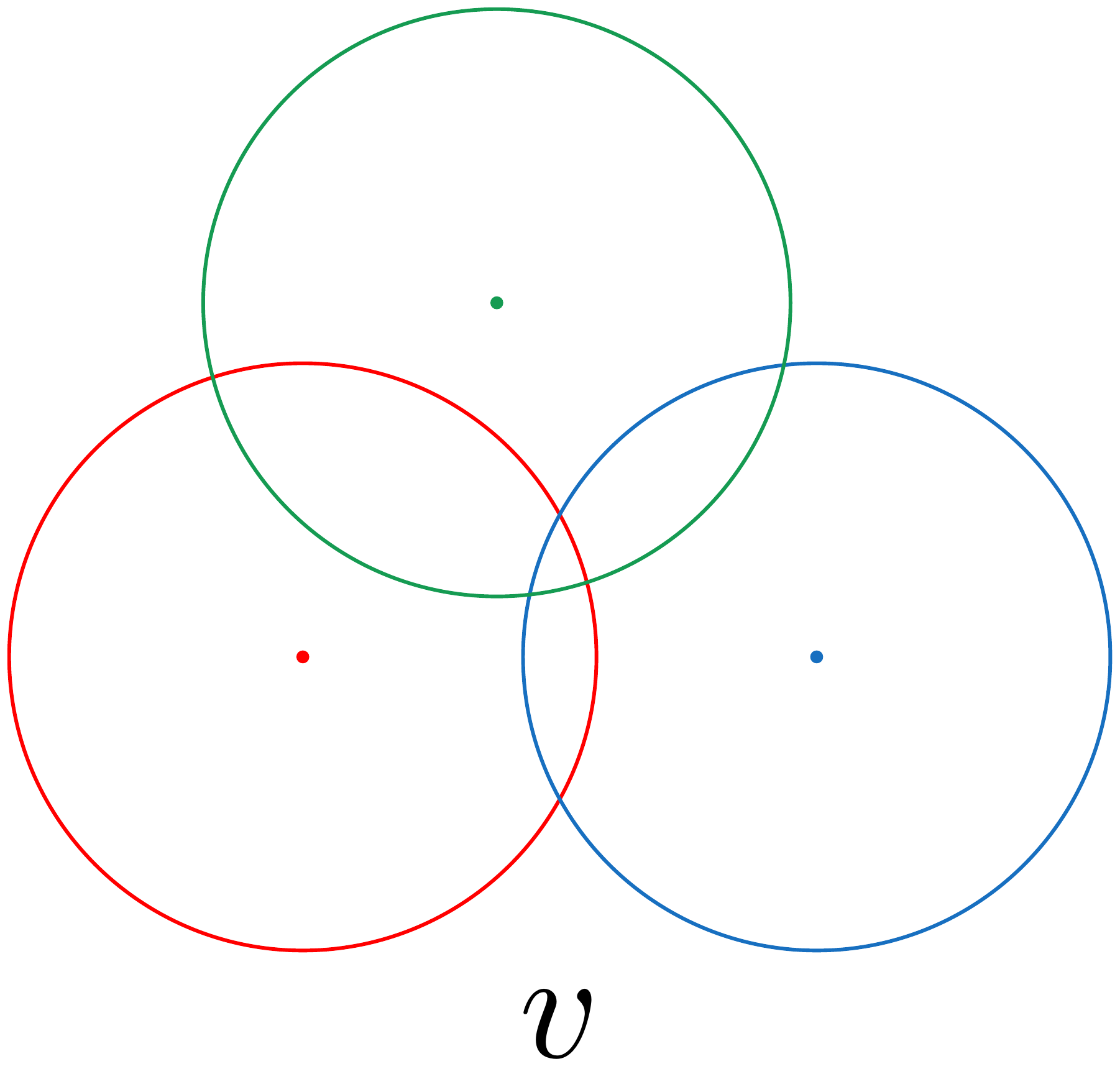}
        \caption*{}
    \end{subfigure}
    
    \caption{The red dashed line represents 8$\pi$ times the Euler charateristic, and the black line represents the third derivative of the volume within the union of spheres. The five different topologies of the union of spheres are shown, along with their respective Euler characteristics and third derivatives of volume plotted.}
    \label{fig:intersections}
\end{figure*}

For a compact convex body $\mathcal{K}$ in the Euclidean $n$-dimensional space $\mathbb{R}^n$, we can draw a surface that is at a distance of $\epsilon$ from the surface of $\mathcal{K}$. The volume within this "$\epsilon$-surface" is (Steiner's formula)
\begin{equation}
    V_{\epsilon} (\mathcal{K}) = \sum_{j = 0}^{n} \binom{n}{j} W_{j}(\mathcal{K}) \epsilon^j
    \label{eq:mink_definition}
\end{equation}
where $W_j$'s are the Minkowski Functionals of $\mathcal{K}$ for $j=0,1,2,...,n$. We will mostly be dealing with $n=3$. Working in 3 dimensions, we can simplify equation~\ref{eq:mink_definition} to get 
\begin{equation}
    V_{\epsilon} (\mathcal{K}) = W_0(\mathcal{K}) + 3W_1(\mathcal{K})\epsilon + 3W_2(\mathcal{K})\epsilon^2 + W_3(\mathcal{K})\epsilon^3
\end{equation}
Thus, taking the appropriate derivatives we get 
\begin{equation}
\begin{aligned}
    V_{\epsilon}(\mathcal{K})\big|_{\epsilon=0} &= W_0(\mathcal{K})\\
    \frac{dV_{\epsilon}(\mathcal{K})} {d\epsilon}\bigg|_{\epsilon=0} &= 3W_1(\mathcal{K}) \\
    \frac{d^2V_{\epsilon}(\mathcal{K})} {d\epsilon^2}\bigg|_{\epsilon=0} &= 6W_2(\mathcal{K})\\ 
    \frac{d^3V_{\epsilon}(\mathcal{K})} {d\epsilon^3}\bigg|_{\epsilon=0} &= 6W_3(\mathcal{K})
    \label{eq:derivatives}
\end{aligned}
\end{equation}
which can be summarized (for $n$ dimensions) as
\begin{equation}
    \frac{d^mV_{\epsilon}(\mathcal{K})} {d\epsilon^m}\bigg|_{\epsilon=0} = \frac{n!}{(n-m)!} W_m(\mathcal{K}) 
    \quad \mathrm{for\hspace{0.1cm}}m=0,1,...,n
\end{equation}
where the prefactor $\frac{n!}{(n-m)!}$ comes from multiplying the coefficient $\binom{n}{m}$ with another $m!$ by differentiating the $\epsilon^m$ term.
\\ \\
To compute the germ-grain Minkowski Functionals, it is essential to evaluate these Functionals not only for individual spheres but also for their unions. Since the unions of spheres may not necessarily be convex, calculating the Minkowski Functionals requires us to have the ability to handle both convex bodies and non-convex unions of multiple convex bodies. We can then use the \textit{additivity} property of Minkowski Functionals. For convex bodies $\mathcal{K}_1,\mathcal{K}_2$: 
\begin{equation}
W_\alpha(\mathcal{K}_1 \cup \mathcal{K}_2) = W_\alpha (\mathcal{K}_1) + W_\alpha (\mathcal{K}_2) - W_\alpha (\mathcal{K}_1 \cap \mathcal{K}_2)
    \label{eq:add}
\end{equation}
We explain the difference between the derivatives of the CDF and the Minkowski Functionals by visualizing what volume is being differentiated in each case. For the Minkowski Functionals of the union of two spheres, we need to use the additivity property (equation~\ref{eq:add}). For two spheres $S_1$ and $S_2$, we can easily calculate $W(S_1), W(S_2)$ using the $V_\epsilon$ method (equation \ref{eq:derivatives}). But to calculate $W(S_1 \cap S_2)$, we need to take derivatives with respect to $\epsilon$ of the volume in figure \ref{fig:alens}. But when we take derivatives with respect to the radius of the entire volume $V(r)$, we implicitly take derivatives of the volume in figure \ref{fig:blens}. Due to this small difference, the Minkowski Functionals and derivatives of the $\mathrm{CDF}_{1\mathrm{NN}}$ are not exactly equal. But as we increase the radius the difference between \ref{fig:alens} and \ref{fig:blens} decreases, and  the Minkowski Functionals and derivatives of the $\mathrm{CDF}_{1\mathrm{NN}}$ approach each other. In fact, for $W_0$ and $W_1$ they are exactly equal for all $r$. The error only shows up in order $\epsilon^2$ and $\epsilon^3$ terms. Let us explicitly calculate the $V_\epsilon$ for two spheres that intersect. 
\\ \\
Consider two spheres of radius $r$ whose centers are a distance $d$ from each other. For the spheres to intersect, we need $d<2r$. The volume of the lens formed by $S_1 \cap S_2$ is given by
\begin{equation}
    V_{\mathrm{lens}}(r,d) = \frac{\pi}{12}(2r-d)^2(d+4r) = \frac{\pi}{12} (16r^3 - 12r^2 d+ d^3)
\end{equation}
The $\epsilon$-volume of the lens is formed by two parts of the spherical shells (of outer radius $R+\epsilon$ and inner radius $R$) subtending a solid angle of $2\pi (1-d/2R)$ at the center of the two spheres, and a toroid with a circular sector (of radius $\epsilon$ and half angle of $\arcsin(d/2R)$) cross section. The volume of the two parts of spherical shells together is
\begin{equation}
    V_\mathrm{shell} = \frac{4\pi}{3} \left( 1 - \frac{d}{2r}\right) (\epsilon^3 + 3\epsilon^2 r + 3 \epsilon r^2)
\end{equation}
The centroid of the circular sector with radius $
\epsilon$ and half angle $\theta$ lies $\frac{2\epsilon}{3\theta}\sin\theta$ vertically from the center. Thus, using Pappus's centroid theorem to calculate the volume of the toroid we get
\begin{equation}
    V_\mathrm{toroid} = 2\pi \left( r \arcsin \left( \frac{d}{2r} \right) \sqrt{1-\left( \frac{d}{2r} \right)^2}\epsilon^2 + \frac{d}{3r} \epsilon^3 \right)
    \label{eq:toroid}
\end{equation}
\\The total $\epsilon$-volume for the lens $S_1 \cap S_2$ is thus
\begin{equation}
\begin{aligned}
    V_\epsilon (r) &= V_{\mathrm{lens}}(r,d) + V_\mathrm{shell} + V_\mathrm{toroid} \\
    &= V_{\mathrm{lens}}(r,d) + 2\pi R(2r-d)\epsilon    \\ &+ 2\pi \left(2r-d + r \arcsin \left( \frac{d}{2r} \right) \sqrt{1-\left( \frac{d}{2r} \right)^2 } \right)\epsilon^2 + \frac{4\pi}{3}\epsilon^3
    \label{eq:v_r}
\end{aligned}
\end{equation}
This $V_\epsilon$ is the volume within figure \ref{fig:alens}. Now let us compute the volume within figure \ref{fig:blens}. It is simply  $V_{\mathrm{lens}}(r+\epsilon,d)$.
\begin{equation}
    V_{\mathrm{lens}}(r+\epsilon,d) = V_{\mathrm{lens}}(r,d) + 2\pi r(2r-d)\epsilon + 2\pi \left(2r - \frac{d}{2}\right)\epsilon^2 + \frac{4\pi}{3}\epsilon^3
    \label{eq:v_rpluspes}
\end{equation}
For $S_1 \cup S_2$, we can write that
\begin{equation}
\begin{aligned}
    \frac{d^mV(r)}{dr^m} &= \frac{d^m}{dr^m} \left( V_{S_1} + V_{S_2} - V_{S_1 \cap S_2} \right) \\
    &= \frac{d^m}{d\epsilon^m} \left( 2V_{\mathrm{sphere}}(r+\epsilon) - V_{\mathrm{lens}} (r+\epsilon,d)\right)\Bigg|_{\epsilon=0}
\end{aligned}
\label{eq:rtoeps}
\end{equation}
Note that the $\epsilon^0$ and $\epsilon^1$ terms in equations \ref{eq:v_r} and \ref{eq:v_rpluspes} are the same. This is expected, as we proved in appendix \ref{sec:arc} that the first two Minkowski Functionals are proportional to the 1NN CDF and first derivative. Even the $\epsilon^3$ is the same in both, but that is not the case when three or more spheres intersect. As $r>>d$, even the $\epsilon^2$ terms approach each other. This means that in equation \ref{eq:rtoeps} the term $V_{\mathrm{lens}} (r+\epsilon,d)$ will start approaching $V_\epsilon (r)$, and thus the Minkowski Functionals and derivatives of the 1NN CDF asymptote to the same value for large radius. This can be geometrically seen in figure \ref{fig:error}.
\\ \\
The error region only contributes in order $\epsilon^2$ for a lens, so even the Euler characteristic and third derivative match each other exactly for the case of two sphere intersections. But for intersection of three or more spheres, there are contributions to the error term in order $\epsilon^3$ also. These errors become vanishingly small as we increase the radius. This can be seen from the decreasing nature of the red region in figure \ref{fig:error}, which starts out being a considerable fraction of $V(R+\epsilon)-V(R)$ (\ref{fig:a}), but shrinks to zero as we keep increasing the radius (\ref{fig:b}). Thus, 
\begin{equation}
    \frac{d^3V(r)}{dr^3} \rightarrow 8\pi\chi(\mathcal{S})
    \label{eq:asymptote}
\end{equation}
where $\mathcal{S}$ is the union of spheres formed at $r$. Since $\chi$ takes only constant values between intersections, and both $\frac{d^3V(r)}{dr^3}$ and $8\pi\chi$ have discontinuities at intersections (see figure~\ref{fig:intersections}), we can use the value of $\frac{d^3V(r)}{dr^3}$ to compute $\chi$ as a function of $r$. To extract the Euler characteristic from these asymptoting functions, we need to be able to find the constant value the function is asymptoting to. 
\\ \\
For the intersection of two spheres, $\frac{d^3V(r)}{dr^3}$ exactly follows $8\pi\chi$. But for three or more spheres intersecting at some $r$, we observe a divergence in $\frac{d^3V(r)}{dr^3}$ at that $r$, followed by a smooth and asymptotic approach to $8\pi\chi$. The divergences occur when either spheres merge, create or destroy a hole or void. Thus, the $r$ at which these divergences happen are either half the distance between two data points or the circumcenter of the triangle formed by three data points. Between intersections, $\frac{d^3V(r)}{dr^3}$ is smooth and infinitely differentiable. Thus, we can fit a Laurent series of the form (for $a<r<b$):
\begin{equation}
    \frac{d^3V(r)}{dr^3} = 
    c_0 + \frac{c_1}{(r-a)} + \frac{c_2}{(r-a)^2} + \frac{c_3}{(r-a)^3} + \cdots
    \label{eq:laurent}
\end{equation}
where $a$ is the $r$ at which $\frac{d^3V(r)}{dr^3}$ diverges, and $b$ is the next divergence length. We do not include $(r-a)^n$ terms because we need an asymptotic approach to $8\pi\chi$ and $(r-a)^n$ will blow up at large $r$. Once we have obtained the coefficients, $c_0$ directly gives $8\pi\chi$ in the range $a<r<b$. Since both $\frac{d^3V(r)}{dr^3}$ and $\chi$ are additive, this method will theoretically work even when there are multiple points and spheres. It is not practical though, because in reality we don't have infinite precision and for a dense set of data points the intersections will happen at very close $r$'s, which will add to the noise. However, theoretically both contain the same information. 

\newpage
\section{Correlation Matrices}
\label{sec:cov}

\begin{figure*}
    \centering
    \includegraphics[width=\textwidth]{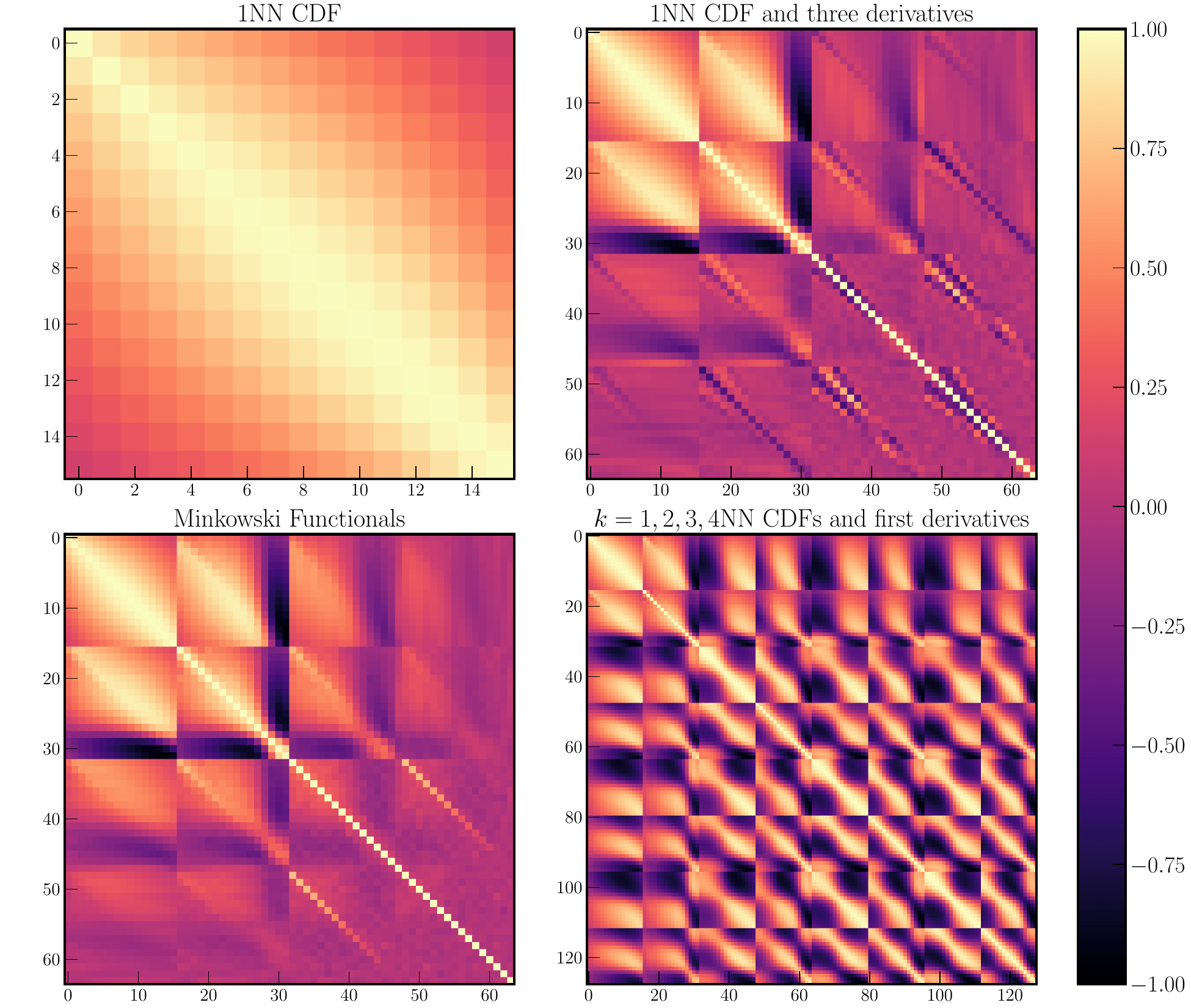}
    \caption{The correlation matrices for a few data vectors. The top left panel is the correlation matrix for the 1NN CDF, the top right panel is for the 1NN CDF and its three derivatives, the bottom left panel is for the four Minkowski Functionals, and the bottom right panel is for $k=1,2,3,4$ NN CDFs along with their first derivatives (each CDF is followed by its first derivative in this data vector). The discontinuous edges in the matrices show where the data vectors have been joined. For example, in the bottom left, the $64 \times  64$ grid can be seen as a block matrix: a $4 \times 4$ grid of $16 \times 16$ sized blocks. Each block represents the correlation between a pair of Minkowski Functionals $\{W_i, W_j\}$, and the discontinuous edges are where two $W_i$'s are joined in the data vector.}
    \label{fig:correlation}
\end{figure*}

The correlation matrix $\mathbf{c}$ can be computed from the elements of the covariance matrix $C_{ij}$:
\begin{equation} 
    \mathbf{c}_{ij} = \frac{ \mathbf{C}_{ij}}{\sqrt{ \mathbf{C}_{ii} \mathbf{C}_{jj}}}
\end{equation}
This ensures that the entries $\mathbf{c}_{ij}$ are within $-1$ and $1$, which allows for the neat visualization of the covariance in the data vectors using the correlation matrix. 
\\ \\
The correlation matrices for some of the data vectors have been plotted in figure \ref{fig:correlation}. The values of $x$ and $y$ in the plots correspond to the entries in the data vector. The top left panel shows the correlation matrix for $\mathrm{CDF_{1NN}}$. The top right panel shows the correlation matrix for the data vector: $\{\mathrm{CDF_{1NN}},\; \frac{d}{dr}\mathrm{CDF_{1NN}},\; \frac{d^2}{dr^2}\mathrm{CDF_{1NN}},\; \frac{d^3}{dr^3}\mathrm{CDF_{1NN}} \}$. The four demarcations within the plot represent where the CDF and derivatives have been joined to form a single data vector. The bottom left panel shows the correlation matrix for the data vector: $\{W_0, \; W_1, \; W_2, \; W_3\}$. The four demarcations within the plot represent where the $W_i$'s have been joined to form a single data vector. The bottom right panel shows the correlation matrix for the data vector : $\{ \mathrm{CDF_{1NN}},\; \frac{d}{dr}\mathrm{CDF_{1NN}},\; \mathrm{CDF_{2NN}},\; \frac{d}{dr}\mathrm{CDF_{2NN}}, \; \mathrm{CDF_{3NN}},$

$\frac{d}{dr}\mathrm{CDF_{3NN}},\; \mathrm{CDF_{4NN}},\; \frac{d}{dr}\mathrm{CDF_{4NN}} \}$. The eight demarcations within the plot represent where the CDFs and their derivatives have been joined to form a single data vector.
\\ \\
As can be seen from the top left correlation matrix, there exists significant correlation between the different bins of the 1NN CDF. The same can be seen from the bottom right matrix for $k=2,3,4$ NN CDFs as well. This is expected, as the CDFs are, by definition, cumulative in nature. So, fluctuations in the CDF at some particular radial bin accumulate and are counted in subsequent bins at larger radii. Since $W_0$ is proportional to $\mathrm{CDF_{1NN}}$, the correlation pattern for $W_0$ resembles that for $\mathrm{CDF_{1NN}}$.
\\ \\
For the derivatives the information in each radial bin becomes more local, as derivatives are sensitive to the local changes of the function. In addition, every successive derivative is washed in more noise. From the first derivatives of the $k$NN CDFs, we see some amount of correlation between radial bins. However, there is negligible correlation between radial bins for the second and third derivatives. We see a similar trend in the Minkowski Functionals as well, with $W_1$ showing some correlation among radial bins, and the correlation diminishes in $W_2$ and vanishes in $W_3$.

\bsp	
\label{lastpage}
\end{document}